\documentclass[12pt,a4paper,oneside]{article}
\usepackage[utf8]{inputenc}

\IfFileExists{lmodern.sty}{\usepackage{lmodern}\def\OFTPClmodern{}}{}

\usepackage[T1]{fontenc}

\usepackage{comment}
\usepackage{xcolor}

\usepackage{lineno}

\usepackage{xspace}

\usepackage{jheppub} 
\usepackage{siunitx}
\usepackage{booktabs} 
\usepackage{amsmath}
\usepackage{braket}
\usepackage{multirow}
\IfFileExists{ulem.sty}{\usepackage[normalem]{ulem}}{}
\usepackage{slashed}

\usepackage{float}
\usepackage{enumitem}

\usepackage{amsthm}
\usepackage{upgreek}
\usepackage{eurosym}
\usepackage{textpos}
\usepackage{array}
\usepackage{subcaption}
\IfFileExists{microtype.sty}{\ifdefined\OFTPClmodern\usepackage{microtype}\fi}{}
\graphicspath{{figures/}}

\def \Fe55		{\ensuremath{^{55}}Fe\xspace}
\def \Be8		{\ensuremath{{^8\mathrm{Be}}}\xspace}

\def \B			{$^{10}\mathrm{B}$\xspace}

\def \B11		{$^{11}\mathrm{B}$\xspace}

\newcommand{\beq}{\begin{equation}}
\newcommand{\bea}{\begin{eqnarray}}
\newcommand{\pair}{\mathrm{e}^{+}\mathrm{e}^{-}}
\usepackage{relsize}
\newcommand{\garfieldpp}{Garfield\nolinebreak[4]\raisebox{.4ex}{\relsize{-3}{\textbf{++}}}\xspace}

\newcommand{\blue}[1]{}
\newcommand{\green}[1]{}
\newcommand{\orange}[1]{}
\newcommand{\red}[1]{}
\newcommand{\magenta}[1]{}
\newcommand{\gray}[1]{}

\mathchardef\mathequals=\mathcode`=
\begingroup\lccode`~=`=
\lowercase{\endgroup\def~}{\mathequals\discretionary{}{\the\textfont0=}{}}
\AtBeginDocument{\mathcode`=="8000 }
\makeatletter
\renewcommand{\neq}{\mathrel{\not\mathrel{\discretionary{}{\the\textfont0=}{}=}}}
\makeatother

\begin{document}


\pagenumbering{arabic}

\title{Track and energy reconstruction algorithms for a time projection chamber with orthogonal fields}
\author[a,b]{Martin Vavřík}
\author[b]{Babar Ali}
\author[b]{Hugo Natal da Luz}
\author[c]{Olivier Rousselle}
\author[b]{Rudolf Sýkora}
\author[a,b]{Tomáš Sýkora}

\affiliation[a]{Institute of Particle and Nuclear Physics,
	Faculty of Mathematics and Physics, Charles University, V Holešovičkách 2, 180 00 Prague 8, Czech Republic}
\affiliation[b]{Institute of Experimental and Applied Physics, Czech Technical University in Prague, Husova 5/240,
110 00 Prague 1, Czech Republic}
\affiliation[c]{Sorbonne Université, Université PSL, 75005 Paris, France}
\emailAdd{tomas.sykora@matfyz.cuni.cz}
\emailAdd{martin.vavrik@utef.cvut.cz}

\date{Prague, \today}

\abstract{%
In this work, we describe the development of track- and energy-reconstruc-tion algorithms for atypical Time Projection Chambers (TPCs) that will be used at the Institute of Experimental and Applied Physics, Czech Technical University in Prague, to search for the anomalous internal pair creation reported by the ATOMKI group. These chambers operate with an inhomogeneous toroidal magnetic field oriented orthogonally to the electric field; we therefore refer to them as Orthogonal-Field TPCs (OFTPCs). Although this configuration distorts the drift of ionization electrons and complicates the resulting electron and positron trajectories, it also offers several practical advantages. We present the most effective of several tested approaches, which employs a simulated ionization-electron drift map for track reconstruction and a Runge--Kutta-based fit for energy reconstruction. Using simulations, we demonstrate that---under idealized conditions, namely an ideal charge readout with no amplification and no noise and with known initial track positions and directions---it is possible to achieve a fitted Gaussian width (sigma) better than 1\% in relative energy for both electrons and positrons, after applying corrections for systematic effects that depend on the track parameters.%
}

\maketitle

\setcounter{tocdepth}{2}


\newpage
\section{Introduction}
In 2015, the ATOMKI group (Hungary) observed in the decay of excited beryllium, produced in the process
\begin{equation}   
    \mathrm{p}+ {^{7}\mathrm{Li}} \rightarrow {^{8}\mathrm{Be}}^{*} \rightarrow {^{8}\mathrm{Be}} + \pair, 
    \label{eq:reaction}
\end{equation} 
an anomaly in the angular distribution of positron-electron ($\pair$) pairs \cite{Krasznahorkay:2015iga} created via the standard Internal Pair Creation (IPC) mechanism \cite{Rose1949}.

Later, a similar anomalous behavior was also observed for other processes in which excited beryllium was replaced by helium {$^{4}$He} nuclei (2019) \cite{Krasznahorkay:2019lyl} and carbon {$^{12}$C} nuclei (2022) \cite{Krasznahorkay:2022pxs}. The observed anomaly has attracted wide attention, as it admits several beyond-standard-model explanations; for recent reviews, see, for example, \cite{atomki_review,Barducci_2023}.
Several groups have announced their interest in similar measurements and are preparing the corresponding detectors \cite{shedding} or have already performed measurements \cite{hanoi,dubna, megii}. Recent results point in opposite directions: the MEG II experiment observed no significant excess in the same $^{7}\mathrm{Li}(\mathrm{p},\pair)\,^{8}\mathrm{Be}$ reaction~\cite{megii}, while the PADME experiment reported a~$\sim\!2\sigma$ excess in resonant $\pair$ annihilation at $\sqrt{s}\approx\qty{16.9}{\MeV}$~\cite{padme_x17}.
Thus, no experiment other than ATOMKI (or linked to the ATOMKI group) has undoubtedly confirmed or disproved the observed anomaly yet. 

\begin{figure}[H]
    \centering
    \includegraphics[width=0.6\textwidth]{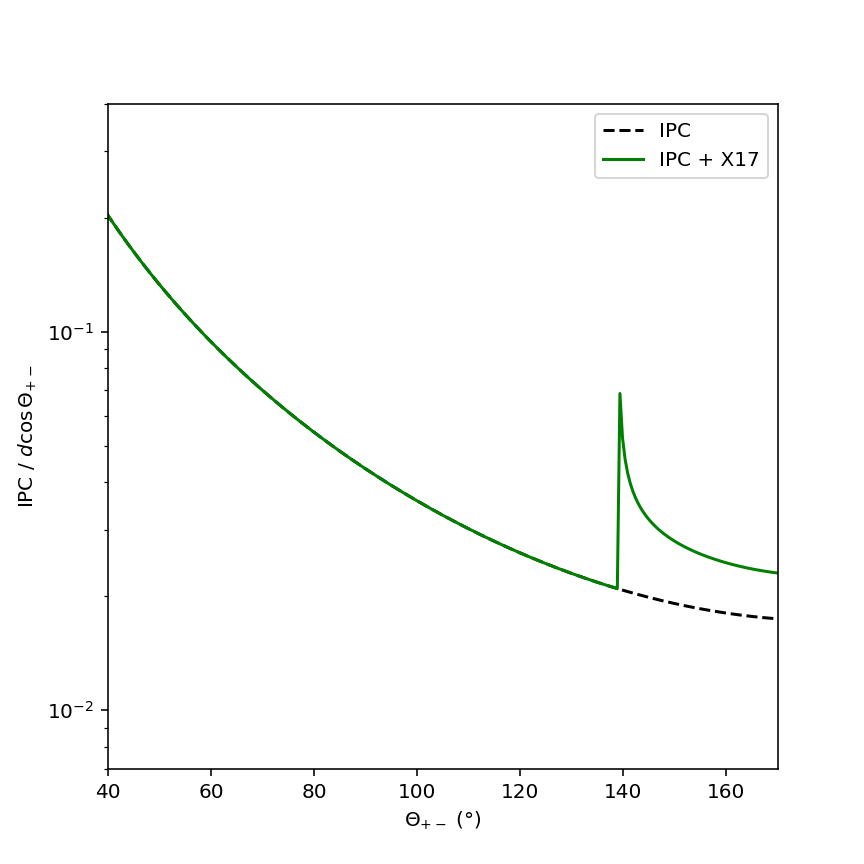}
    \caption{Angle $\theta$ between the constituents of the $\pair$ pairs for the combined IPC and X17 production, obtained from an analytical calculation \cite{Rose1949}. A branching ratio X17/IPC of \num{3E-3} was used.} 
    \label{fig:IPCplusX17_ideal}
 \end{figure}

\begin{figure}[H]
    \centering
    \includegraphics[width=0.7\textwidth]{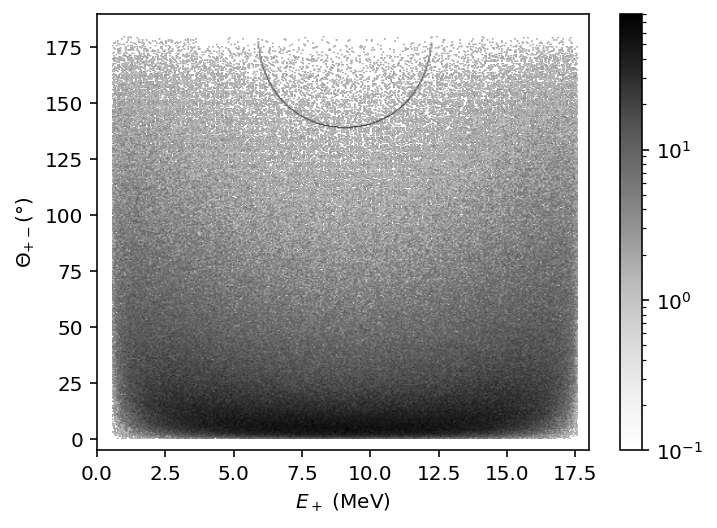}
    \caption{2D histogram of the pair angle vs.\ positron energy, for pairs created in the standard Internal Pair Creation process (here a mixture of transitions M1 + 0.21 E1 is considered) and for pairs created by the decay of a hypothetical boson X17. The gap in the signal pattern from the decay is clearly visible on top of the background. The data used for the plot come from a~simulation; no detector effects affecting the precision of either the angles or the energies are included here, i.e., the true, generated values are plotted. To generate the plot, $10^6$ events were used, which represent an expected lower bound on the number of events to be measured. A branching ratio X17/IPC of \num{3E-3} was used. }
    \label{fig:theta-E-plot}
\end{figure}

The \textit{shape} of the one-dimensional histogram of the distribution of the angular correlation between the directions of the electron and positron
produced by a detector with a very high (infinite) resolution would serve as convincing proof of the existence of a decaying particle, see Fig.~\ref {fig:IPCplusX17_ideal}.
There is, however, no such detector, and the resolutions are limited. 
The finite angular resolution of a real detector smears the peak structure and makes any conclusion about the \textit{origin} of the peak, based on its shape alone, difficult. Alternatively, the two-dimensional histogram correlating the angle $\theta$ between the constituents of the $\pair$ pairs with the positron energy, see Fig.~\ref{fig:theta-E-plot}, can be used, because the X17 decay populates a region that is well separated from the IPC background (the ``separation gap'').

Both resolutions in angle $\theta$ and energy $E$ must be high enough to keep the signal distinguishable from the background.

The present article is devoted to the track and energy reconstruction algorithms for the non-standard Orthogonal-Field Time Projection Chamber (OFTPC) that forms part of the apparatus searching for X17 at the Van de Graaff (VdG) accelerator of the Institute of Experimental and Applied Physics, Czech Technical University in Prague~\cite{VdG}. The angle between the constituents of the pair is currently reconstructed from the Timepix3 detectors~\cite{triangle} (see Fig.~\ref{fig:TPC_schematics_EB}); this measurement is affected by multiple scattering in the vacuum tube and in the Timepix3 detectors themselves, and the corresponding angular resolution is taken from dedicated measurements~\cite{triangle}. The OFTPC could in principle provide a complementary angle measurement, although an angle reconstructed from the OFTPC would be degraded further, since the leptons must first traverse all the preceding material; a comparison of the two methods is planned for the future.

The paper is organized as follows. Section~2 describes the experimental setup and the design of the OFTPC. Section~3 presents the three-step reconstruction pipeline: drift mapping, track simulation, and Runge--Kutta energy fit. Section~4 discusses the results and states the conclusions.

\section{Experimental setup}
\subsection{Use of the VdG accelerator and general concept of the detector}

Excited {$^{8}$Be} nuclei, with excitation energies of 17.64 or \qty{18.15}{\MeV}, are produced in collisions of protons accelerated by the VdG accelerator \cite{VdG} to the corresponding resonance energies of \qty{441}{\keV} or \qty{1.03}{\MeV}, respectively, with a thin disk-like target containing lithium.

\begin{figure}[H]
    \centering
    \includegraphics[width=0.5\textwidth]{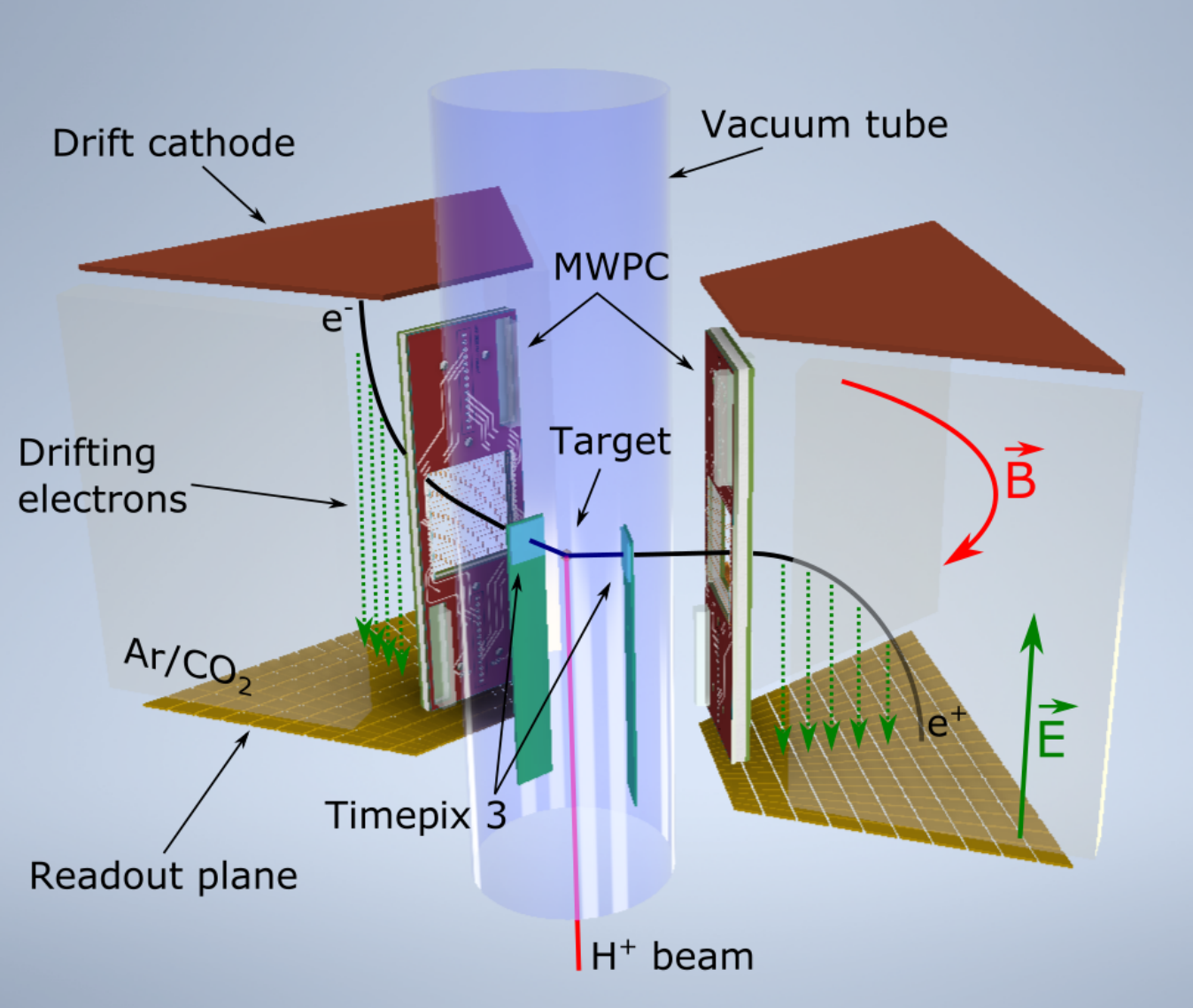}
    \caption{Schematics of the OFTPC design, showing the electric $\vec{E}$ and magnetic $\vec{B}$ fields and three layers of detectors: 1) Timepix3, for the reconstruction of pair origin vertices and particle tracks, 2) multi-wire proportional chambers (MWPC), providing information about the position of charged particles and the direction of their momentum, and 3) Time Projection Chambers, reconstructing the energy of each particle; the readout pad plane is at the bottom here.}
    \label{fig:TPC_schematics_EB}
\end{figure}

Particles originating from the de-excitation of {$^{8}$Be} nuclei and subsequent decays, including $\pair$ pairs, are detected by three layers of detectors surrounding the target, see Fig.~\ref{fig:TPC_schematics_EB}.
The first layer, see Fig.~\ref{fig:TPC_schematics_EB} (where only two of the six sectors are shown), consists of six Timepix3 (TPX3) detectors with their surfaces parallel to the beam and to the sides of a hexagon. These detectors serve to reconstruct the $\pair$ origin vertices and the angle between the constituents of the $\pair$ pairs, for which we use the abbreviation~\(\theta_{+-}\).
\begin{figure}[H]
    \centering
    \begin{subfigure}{0.4 \textwidth}
        \includegraphics[width=\textwidth]{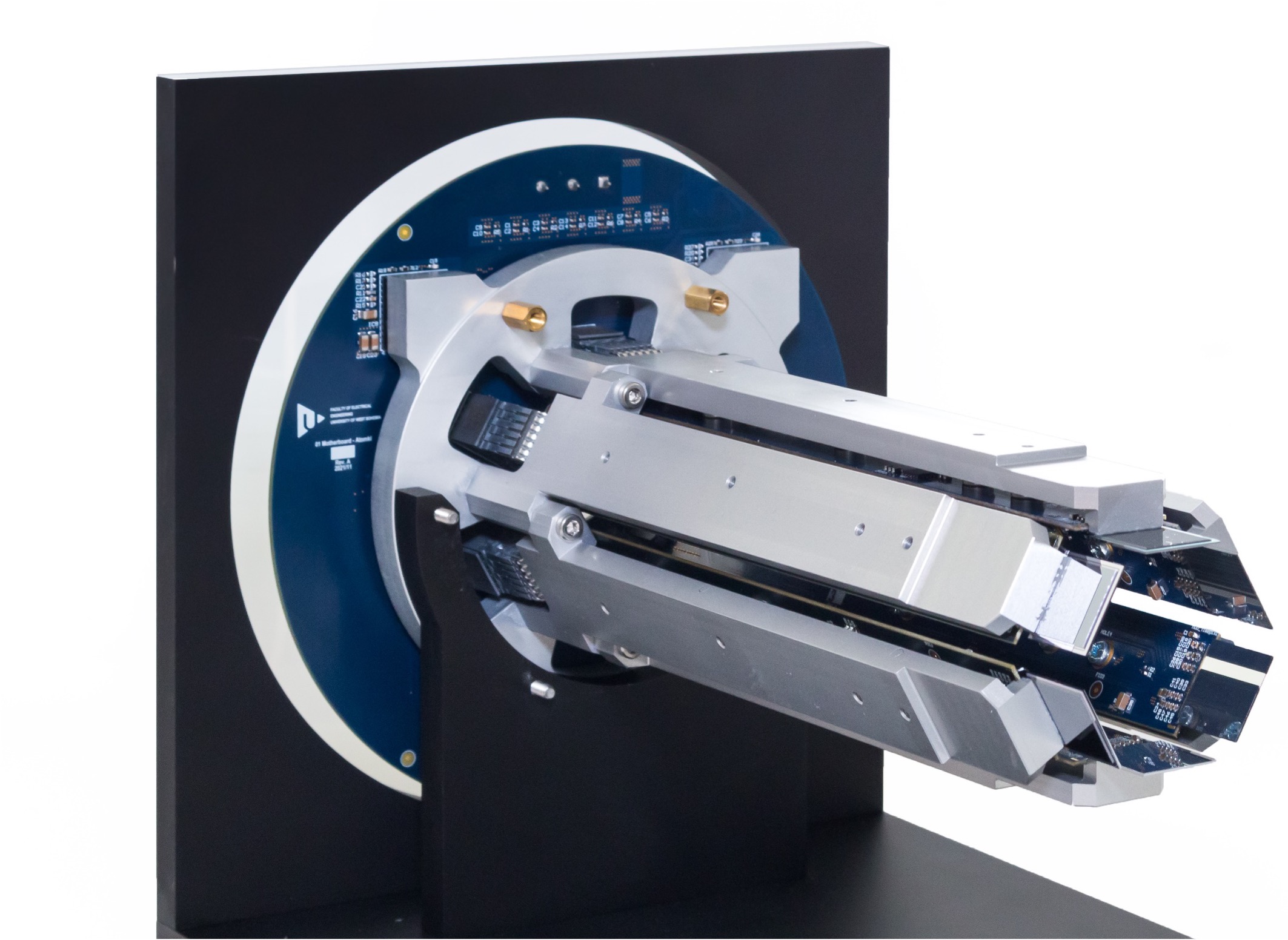}
        \caption{}
        \label{fig:TPX3hexaCAD_a}
    \end{subfigure}
    \hfill
    \begin{subfigure}{0.54\textwidth}
        \includegraphics[width=\textwidth]{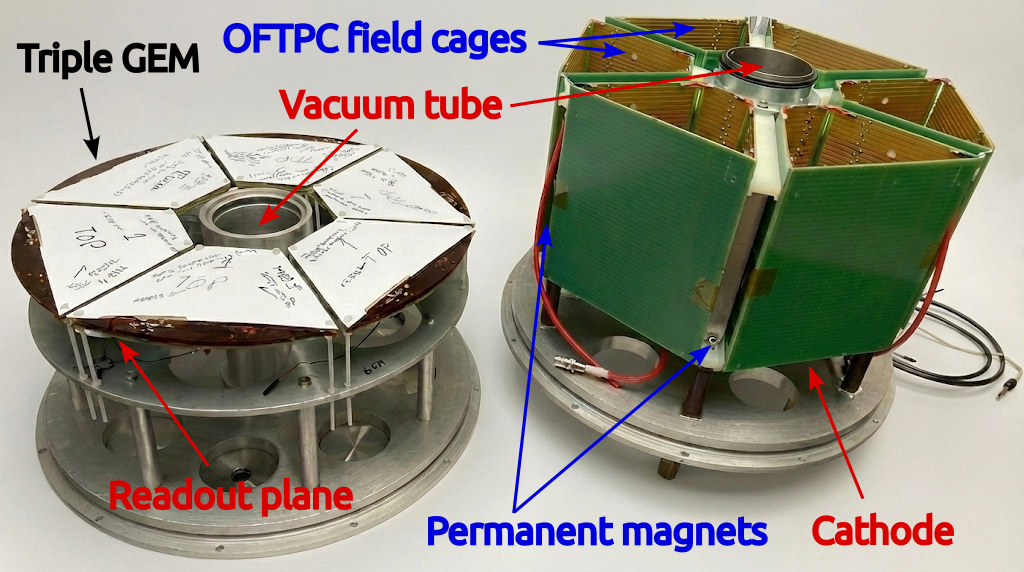}
        \caption{}
        \label{fig:TPX3hexaCAD_b}
    \end{subfigure}
    \caption{Parts of the detector to be mounted at the Van de Graaff accelerator: (a) photograph of the assembled TPX3 detector, (b) a section of the full setup showing the beam (vacuum) pipe (center), the permanent magnets, and the hexagonal mounting frame (originally designed by the ATOMKI group) with readout pad plane based on Gas Electron Multipliers (GEMs).}
    \label{fig:TPX}
\end{figure}
The second layer of detection consists of multi-wire proportional chambers (MWPC). Combining the position of the last hit pixel of the TPX3 with the MWPC position yields the direction information used later in the energy reconstruction. The resulting angular resolution, dominated by multiple Coulomb scattering in the target and detector material, is characterized in~\cite{triangle}. Since the measurement in~\cite{triangle} was performed with a triangular Timepix3 arrangement, the resolution quoted there is used here as a conservative (upper) estimate for the present hexagonal configuration.

Finally, the energy of each particle is reconstructed using the third layer, consisting of the six segments of OFTPCs. Between each segment, strong permanent magnets define a magnetic field that is perpendicular to the electric field. The final design of the TPC system, with orthogonal electric and magnetic field directions, was chosen to keep the construction simple and to minimize the number of electronic channels needed to reach the required energy resolution.
Further technical details can be found in \cite{Vartin_Vit_Vavrik_thesis, triangle}.

\subsection{Time projection chamber with orthogonal fields}
Time projection chambers are used in many high-energy particle physics experiments, see e.g. \cite{Hilke_2010, Blum:2008nqe}. In their standard designs, the electric and magnetic fields are oriented parallel to each other. In such a case, there is no shift of the drift direction (Lorentz angle), the magnetic field may be increased as far as technically feasible, the gas pressure can be kept low, and the transverse diffusion of the electrons drifting in the gas is kept very small (see, e.g., \cite{Blum:2008nqe}, Section 11.7). 

 One of the most important challenges when designing TPCs and reconstructing their data is related to the effect caused by the non-vanishing force in the direction of $\vec{E}\times\vec{B}$, and is usually linked to non-uniformities of the fields that, in some regions of the drift volume, lead to the existence of a~component of the magnetic field orthogonal to the electric field. The resulting force is no longer parallel to the electric field and generates distortions in the reconstructed trajectories. This effect degrades the accuracy of the reconstructed particle energy.\footnote{In a conventional (parallel-field) TPC, the $\vec{E}\times\vec{B}$ correction (see, e.g., the Langevin approximation in~\cite{Hilke_2010}) is negligible in the bulk of the drift volume but becomes significant wherever mechanical or space-charge imperfections create a local component of $\vec{B}$ perpendicular to $\vec{E}$. Because these imperfections are localized and hard to characterize globally, achieving sub-percent accuracy requires extensive field characterization. In the OFTPC the situation is qualitatively different: the large $\vec{E}\times\vec{B}$ deflection exists by design and is globally well-defined. Corrections can therefore be tabulated once, for a fixed detector configuration and field map, over the entire drift volume, through the drift map described in Section~3.}
 However, orthogonal-field designs are not without precedent; see for example the radial-drift TPCs of CERES/NA45~\cite{ceres}, BONuS12~\cite{bonus}, and ALPHA-g~\cite{alpha_rtpc}.

For small TPCs, the distortions caused by the orthogonality of \(\vec{E}\) and \(\vec{B}\) fields are more tractable: the ionization electrons drift over limited distances, so the integrated $\vec{E}\times\vec{B}$ deflection remains moderate and allows straightforward correction. In fact, if the fields are orthogonal by design and the corrections can be defined over the entire volume, the field non-uniformities become less important. In our experiment, permanent magnets are used to generate a toroidal magnetic field. This results in a gas volume that is divided by the magnets into six independent segments. As the $\vec{B}$ field bends the positrons towards the readout plane (and the electrons away from it), their trajectories remain within the volume segment, together with all generated secondary particles. The high energy resolution of the OFTPC is a consequence of the good timing resolution of the readout plane: because the primary particles are bent in the drift (timing) direction, fine timing resolution suffices; if instead they were bent in the plane of the readout, a much finer spatial pixelization would be required. The magnetic deflection of the drifting ionization electrons is, in the end, relatively small and can be corrected for, as demonstrated later in this work. The gas mixture also plays a role: the OFTPC is filled with Ar/CO$_2$ (70:30) at atmospheric pressure, with a drift electric field of around \qty{400}{\volt\per\centi\metre} (corresponding to approximately \qty{6}{\kilo\volt} across the drift volume), a regime in which this mixture is close to its minimum of transverse diffusion~\cite{Peisert:154069}. It can be shown (see e.g.\ Eq.~22 in~\cite{Vartin_Vit_Vavrik_thesis}) that a gas with a lower drift velocity reduces the $\vec{E}\times\vec{B}$ distortion and improves the precision of the $z$-coordinate reconstruction. Permanent magnets have been used to create the magnetic field of a TPC in other experiments as well; for example, the FASER experiment~\cite{faserTDR} uses a permanent magnet with a field of \qty{0.6}{\tesla}. More details concerning the construction of the OFTPC are provided in \cite{Vartin_Vit_Vavrik_thesis, MPGD2021, triangle}.

A further reason for using permanent magnets, beyond their cost and effectiveness, is technical. A small experiment such as this one justifies neither the additional complexity nor the space required to generate the necessary strong magnetic field (\qty{0.3}{\tesla}) with an electromagnet.

When permanent magnets of a simple block shape are used, the simplest solution for a TPC module is shown in Fig.~\ref{fig:TPC_schematics_EB}.
This configuration results in mutually perpendicular electric and magnetic fields throughout most of the volume, requiring only a limited number of corrections at the boundaries of the magnets.

\section{Track and energy reconstruction algorithms for the OFTPC}
In our experiment, aimed at the detection of the hypothetical X17 boson, the IPC background contribution is expected to be dominant, see Fig.~\ref{fig:IPCplusX17_ideal}. Proper reconstruction of the particle energy is therefore essential for isolating a clean sample in the two-dimensional $(\theta_{+-}, E)$ distribution shown in Fig.~\ref{fig:theta-E-plot}.

Particles entering the OFTPC volume are called primary particles; those produced inside it by ionization are called ionization electrons. Primary particles of interest---electrons and positrons from the $\pair$ pair---have kinetic energies in the range \qtyrange{3}{13}{\MeV} (Section~\ref{sec:simulation}), corresponding to Lorentz factors $\gamma \approx 7$--$26$, i.e.\ in the relativistic regime where the specific ionization energy loss is close to its minimum-ionizing value. In the Ar/CO$_2$ (70:30) gas at atmospheric pressure, the mean energy deposited over the full depth of the OFTPC is at most a few tens of keV, negligible compared to the total kinetic energy. The trajectory of each primary particle is therefore governed almost entirely by the Lorentz force, with energy loss causing no measurable change in its deflection. The electric field $\vec{E}$ is considered to be uniform in the OFTPC volume; the non-uniform magnetic field $\vec{B}$ is given by the superposition of the magnetic fields of the permanent magnets. In a first approximation, both fields are considered ideal (as calculated by finite element methods), and their values are defined at the 3D space grid points. Later, the ideal values will be replaced by the measured values.

\begin{figure}[H]
	\centering
	\includegraphics[width=0.7\textwidth]{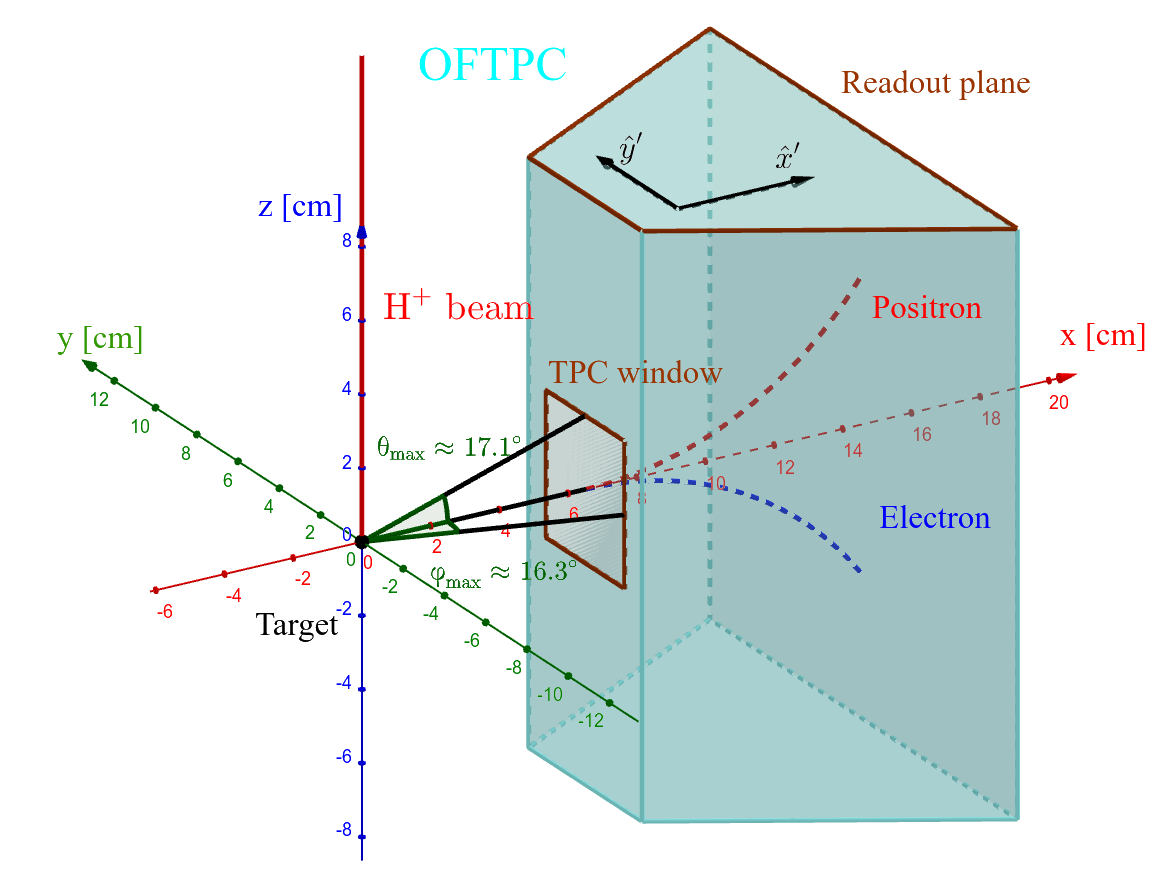}
	\caption{The coordinate systems and the OFTPC detector: ($x,y,z$) coordinates are used for the description of the full detector (and tracks), coordinates ($x',y',t$) for the pad plane. The positions of the beam, target, and window as well as the position and coordinates of the readout plane are shown. The spherical angles $\theta$ and $\varphi$ (measured from the equatorial \textit{xy} plane) are used to describe the particle momentum direction. Made with GeoGebra\textsuperscript{\textregistered}.}
	\label{fig:OFTPC_coordinates}
\end{figure}

\begin{figure}[H]
        \centering
        \includegraphics[width=0.65\textwidth]{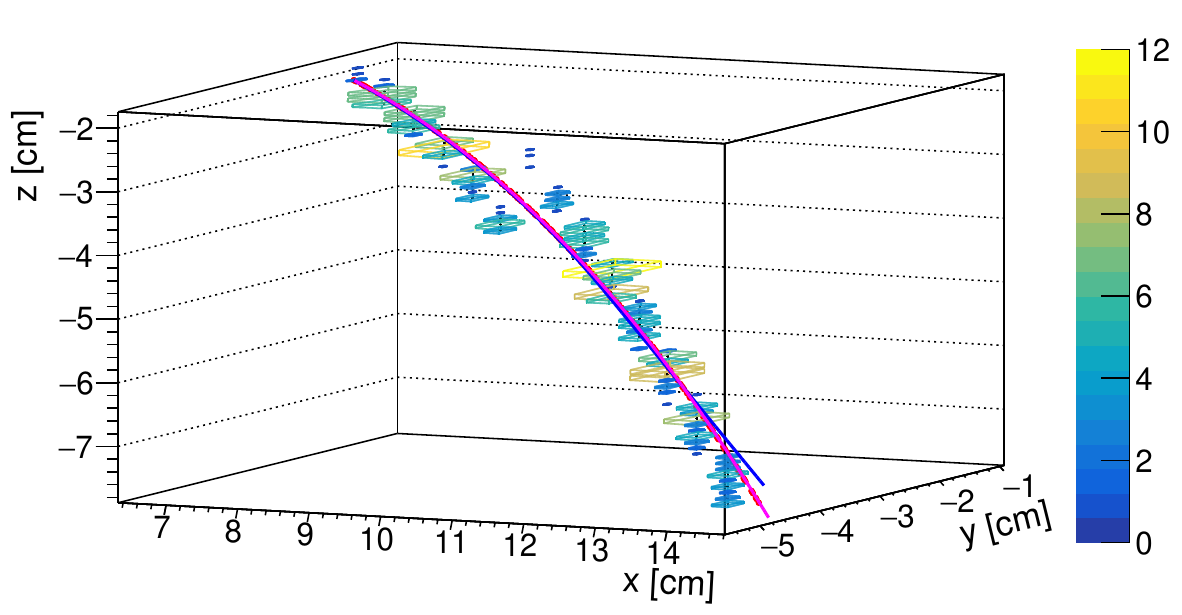}
        \caption{Example of a Runge--Kutta track fit (solid curve) overlaid on the reconstructed voxels (coloured by charge) for a simulated \qty{8}{\MeV} electron. The fitted energy is used as the basis for all energy-reconstruction results presented below.}
        \label{fig:rkfit_example}
    \end{figure}

Energy reconstruction, along with its testing on simulations, requires three steps:
\begin{enumerate}
    \item \textbf{Mapping of the drift:} In a~TPC with an ideal (continuous) readout, each ionization electron produced by an interaction of the primary particle at the coordinates $(x,y,z)$ is registered on the readout as $(x',y',t)$, see Fig.~\ref{fig:OFTPC_coordinates}. Since the drift path of the ionization electrons is distorted by the $\vec{E}\times \vec{B}$ effect, we cannot simply assume a direct correspondence between these coordinates, i.e., $(x,y,z) \neq (x',y',v_\text{d} t)$, where $v_\text{d}$ is the drift velocity. Therefore, a~mapping~$\mathcal{M}$ between the two spaces must be determined. To get an approximation of the mapping, the drift of a large number of electrons from known initial positions covering the detector volume towards the readout was simulated. The process of charge multiplication via a Gas Electron Multiplier (GEM; located at the position of the readout plane) is neglected here.
    \item \textbf{Simulation and reconstruction of primary tracks:} Primary electron and positron tracks entering the detector were simulated, as well as the drift of the ionization electrons generated by the interactions with the gas. The positions of the interactions along the track were reconstructed in the form of track voxels from the detector data (charge $q(x_\text{pad},y_\text{pad},t)$ in each pad and time bin) using the map determined in the previous step.
    \item \textbf{Energy reconstruction:} Based on the reconstructed voxels with charge information, the particle track is fitted with the kinetic energy as the free parameter. An initial estimate is obtained from a circular prefit to the projected voxels, providing a seed for the MIGRAD minimization. An example of the final Runge--Kutta fit overlaid on the reconstructed voxels is shown in Fig.~\ref{fig:rkfit_example}.
\end{enumerate}

The MWPC and TPX3 detector layers will provide boundary conditions for each track: the position of a charged particle entering the volume of the OFTPC will be measured by the MWPC (with a precision of \qty{500}{\um}), and the direction of the particle's momentum at its entrance to the OFTPC can be estimated by combining the last TPX3 pixel (size \qtyproduct{55x55}{\um}) with the position measured by the MWPC. In this paper, we assume that these conditions are known exactly; in the future, they may be further refined in a fit.

\subsection{Mapping of the drift}
To account for the drift distortion by the magnetic field, a simulation is needed for the reconstruction procedure --- the first step is to get a (direct) mapping from the detector space $\mathcal{D}$ to the readout space $\mathcal{R}$
    \begin{equation}
        \overline{\mathcal{M}}\colon \mathcal{D} \rightarrow \mathcal{R},\ (x,y,z) \mapsto (\overline{x}',\overline{y}',\overline{t})
        \label{average_mapping}
    \end{equation}
of the average readout coordinates (position on the readout plane $\overline{x}',\overline{y}'$ and drift time~$\overline{t}$) corresponding to the ionization vertices $(x,y,z)$ inside the OFTPC.

For the drift movement of ionization electrons in the OFTPC, we use the \garfieldpp toolkit~\cite{Garfield++}; the subsequent development of avalanches in the GEM foils could in principle be added in the future, but is not simulated here. \garfieldpp is a toolkit for the detailed simulation of gaseous and semiconductor ionization detectors. There are two basic options for ionization electrons in \garfieldpp:
\begin{enumerate}
    \item \texttt{AvalancheMC} (Avalanche Monte Carlo) tracking, which requires a gas table calculation and applies a random diffusion step using precalculated coefficients,
    \item \texttt{AvalancheMicroscopic} tracking, which uses the equation of motion to follow electrons from collision to collision and models scattering.
\end{enumerate}
The latter is generally slower but more precise, especially for small structures. It was used in this work because the magnetic field produces tilted, curved drift trajectories (see Fig.~\ref{fig:TPC_diffusion_drift}), and microscopic tracking captures these in full detail.

\begin{figure}[hbtp]
        \centering
        \includegraphics[width=0.45\textwidth]{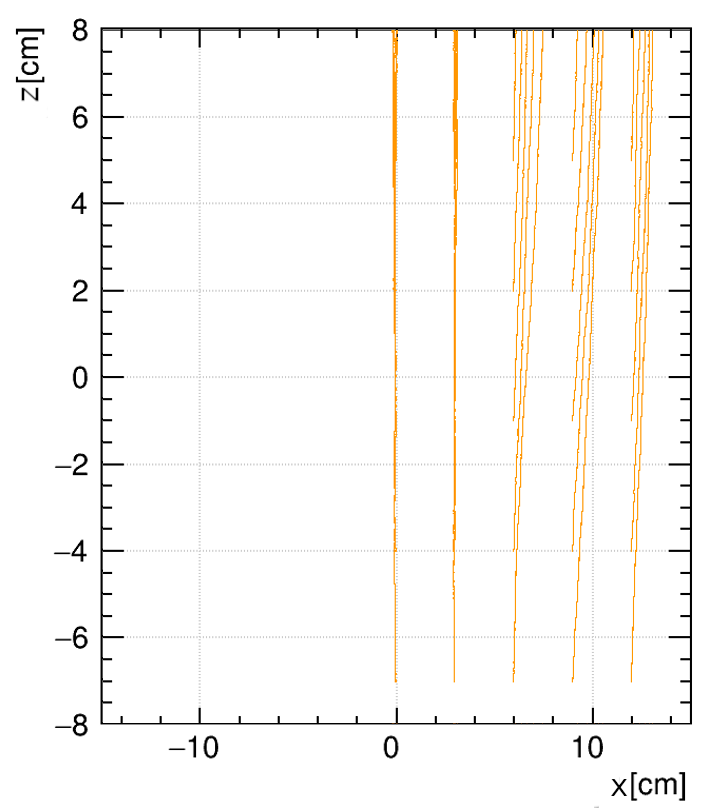}
        \caption{An example of a simulation of the mapping $\mathcal{M}$, showing how regular grid $(x,y,z)$ points are mapped to readout plane coordinates and providing a base for understanding the resulting irregularity of $(x',y',t)$ grid points.}
        \label{fig:TPC_mapping_direct}
    \end{figure}

The (direct) mapping is created using the microscopic simulation of the motion of ionization electrons. The initial positions of these electrons are placed on a regular (Cartesian) grid in detector space $(x,y,z)$. Each electron is initialized with a kinetic energy of 0.1 eV (\garfieldpp does not accept a zero initial energy) and an isotropic random direction. This energy is higher than the mean thermal energy at room temperature ($k_\mathrm{B}T \approx 25$ meV); the corresponding speed ($\sim\!\qty{2e5}{\metre\per\second}$) is far smaller than the instantaneous electron speed between collisions in the drift field, but much larger than the macroscopic drift velocity ($\sim\!\qty{e4}{\metre\per\second}$). We simulate $n = 100$ ionization electrons at each grid point and average their readout coordinates:
    \begin{equation}
        \mathcal{\overline M}\colon \mathcal{D} \rightarrow \mathcal{R},\ (x,y,z) \mapsto \frac{1}{n}\sum_{i=1}^{n}{(x'_i, y'_i, t_i)}.
    \label{inverse_mapping}
    \end{equation}
The choice $n = 100$ is a practical compromise. With 100 independent realizations per grid point the statistical uncertainty on the mean landing position is reduced to $\sigma_\text{diff}/\!\sqrt{100}$, where $\sigma_\text{diff}$ is the single-electron diffusion spread. Since diffusion itself is the dominant source of residuals in the reconstructed track (see Fig.~\ref{fig:residues_hist}), reducing the map's statistical uncertainty below $10\%$ of $\sigma_\text{diff}$ adds negligible benefit. For a visualization of such a simulation, see Fig.~\ref{fig:TPC_mapping_direct}. We also obtain information about the distribution of these readout positions due to the randomness of electron collisions with gas particles.\footnote{From preliminary tests (e.g., Mardia's test of skewness and kurtosis \cite{mardia}), we see that the distribution of readout positions of ionization electrons caused by diffusion is well-described by a multivariate normal distribution. This information can be used to speed up the track simulation with Monte Carlo.}

    \begin{figure}
    	\centering	\includegraphics[width=1.\textwidth]{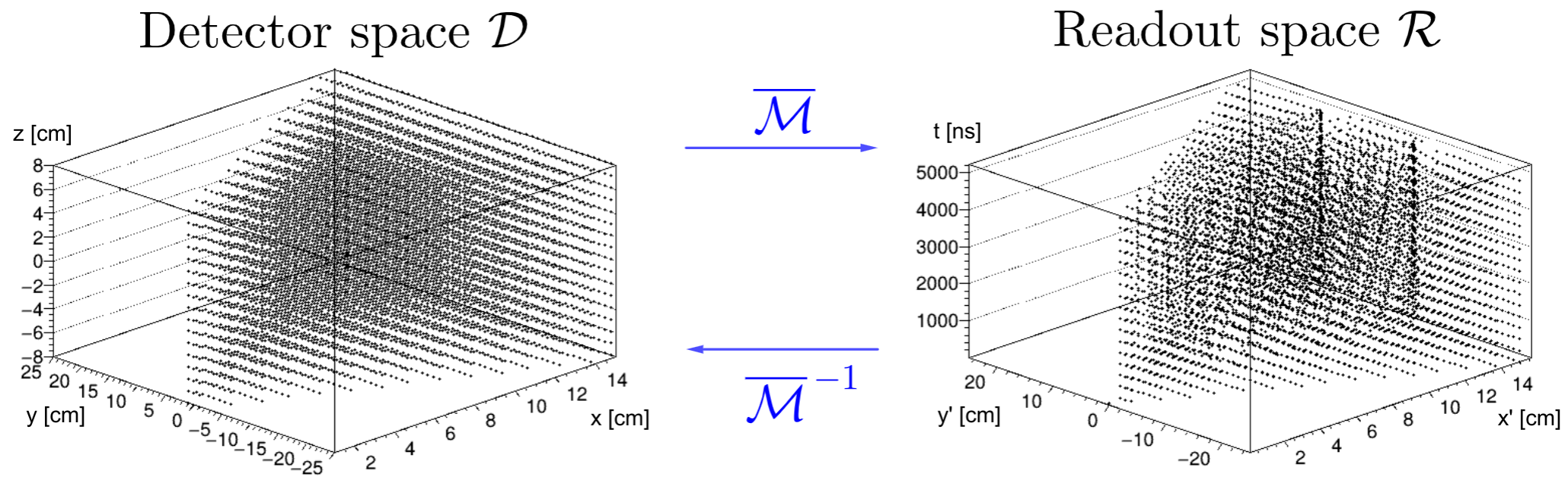}
    	\caption{A visualization of the direct mapping $\overline{\mathcal{M}}$ between the detector coordinate space and the corresponding readout coordinate space and of its inverse $\overline{\mathcal{M}}^{-1}$.}
    	\label{fig:TPC_mapping_visualization}
    \end{figure}

\begin{figure}
    \centering
    \begin{subfigure}{0.45\textwidth}
        \centering
        \includegraphics[width=\textwidth]{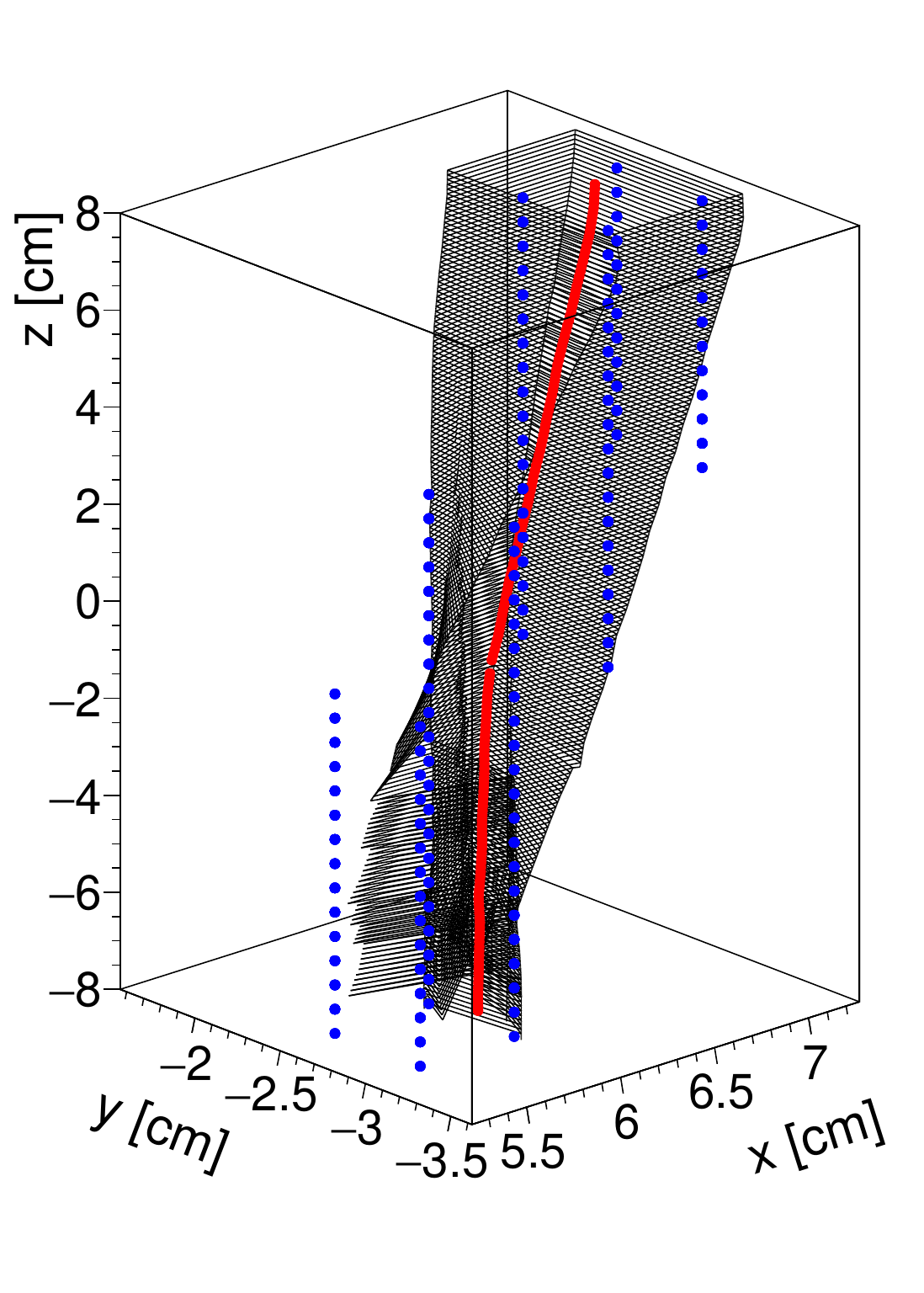}
        \caption{Polynomial interpolation.}
    \end{subfigure}
    \begin{subfigure}{0.45\textwidth}
        \centering
        \includegraphics[width=\textwidth]{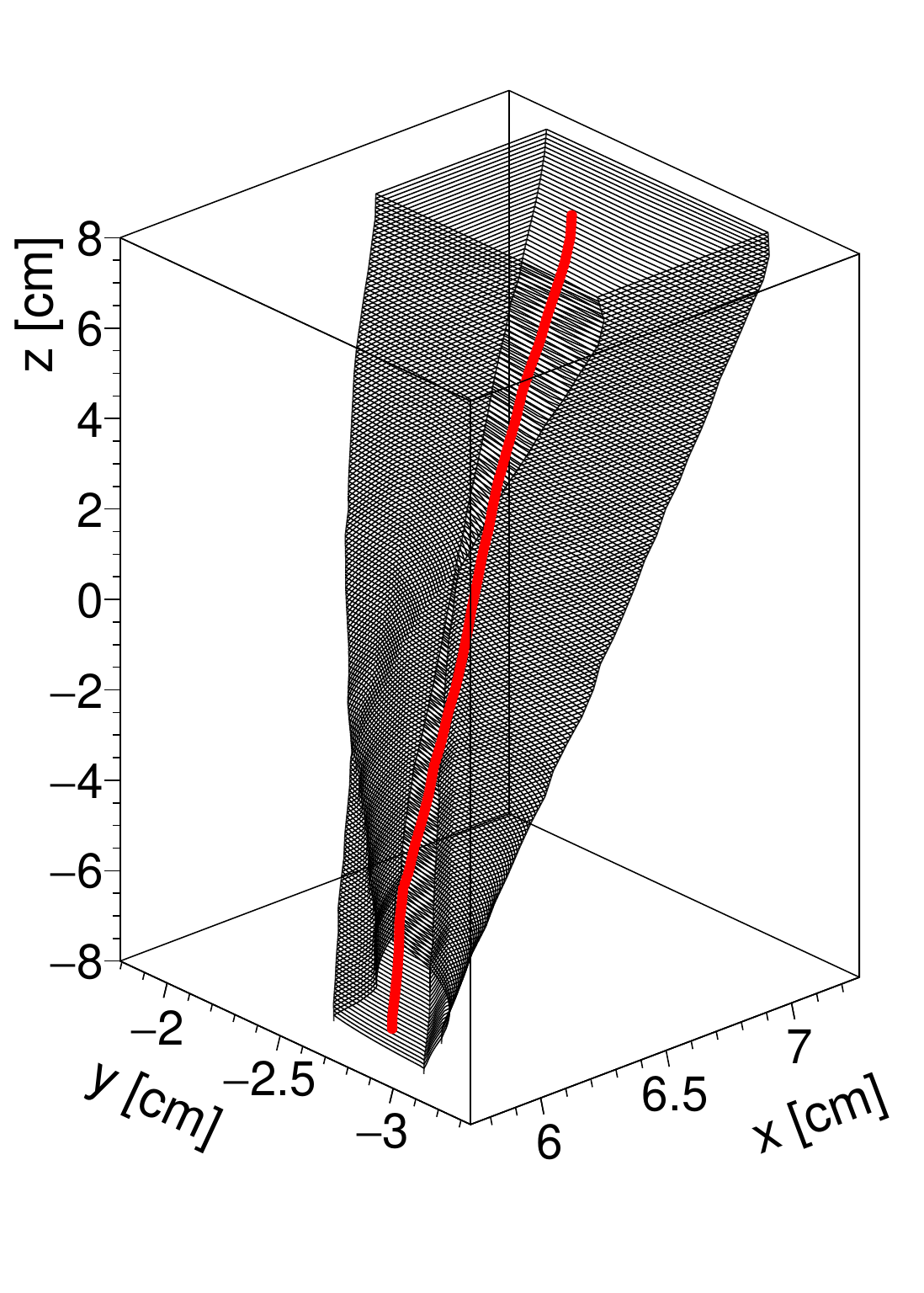}
        \caption{Gradient descent.}
    \end{subfigure}
    \caption{Comparison of the reconstruction of boundaries of the problematic pad 12 near the magnet pole for different drift times (time step \qty{100}{\ns}) using the two methods of $\overline{\mathcal{M}}$ inversion. The reconstructed centers of the pad and time bin are denoted by red points, and the simulated map points used for the interpolation are in blue. Polynomial interpolation in this region with high distortion leads to unwanted artifacts for large drift paths (spikes in the black boundaries, less significant effect on the red centers). Only ten corner pads are affected (8 of them near the vacuum tube), and in current simulations, electrons in these regions are very rare.}
    \label{fig:pad12}
\end{figure}

Between grid points, the direct map~$\overline{\mathcal{M}}$ is evaluated by trilinear interpolation on the regular detector-space grid; this forward interpolant serves only to speed up the simulation of primary tracks (Section~\ref{sec:simulation}), the reconstruction itself relying on the inverse map~$\overline{\mathcal{M}}^{-1}$ described next. The inverse map $\overline{\mathcal{M}}^{-1}$ is built from the exact map at the grid points together with an interpolation between them. Because the forward and inverse directions use different interpolation schemes, $\overline{\mathcal{M}}$ and $\overline{\mathcal{M}}^{-1}$ are not exact inverses of each other away from the simulated grid points. This mismatch is negligible in the bulk of the volume; it becomes relevant only for the faster (polynomial) inversion method in the high-distortion regions near the magnet poles, where its results can cease to be reliable (see Fig.~\ref{fig:pad12}).

We tested two methods of inversion --- polynomial interpolation on the irregular grid in the readout space (the main, faster method), and trilinear interpolation on the regular grid combined with a gradient-descent search for the inverse (slower, but better in some extreme cases near the poles of the magnets, see Fig.~\ref{fig:pad12}). In the first method, we use a binary search algorithm \cite{enwiki:binary_search_alg} to find the eight grid points that surround\footnote{This is non-trivial, since these points alone do not define a closed volume; instead, we require that each coordinate of the target point lie between the minimum and maximum of the corresponding coordinates of the eight points.} the point we want to map, effectively forming a pseudocube around it. Due to the uneven spacing of these points (see the right part of Fig.~\ref{fig:TPC_mapping_visualization} with the resulting grid in the readout coordinate system), we can no longer use trilinear interpolation to find the resulting point in the detector space. Instead, we use a similar interpolation method, which includes the determination of the coefficients of a~polynomial
    \[
        f(x',y',t) = ax'y't+bx'y'+cx't+dy't+ex'+fy'+gt+h
    \]
for each of the reconstructed coordinates $\tilde{x}, \tilde{y}, \tilde{z}$ in the detector space based on the surrounding points. The results presented in this article use the first method only.

\subsection{Simulation and reconstruction of primary tracks}\label{sec:simulation}
\begin{figure}[H]
    \centering
    \includegraphics[width=0.6\textwidth]{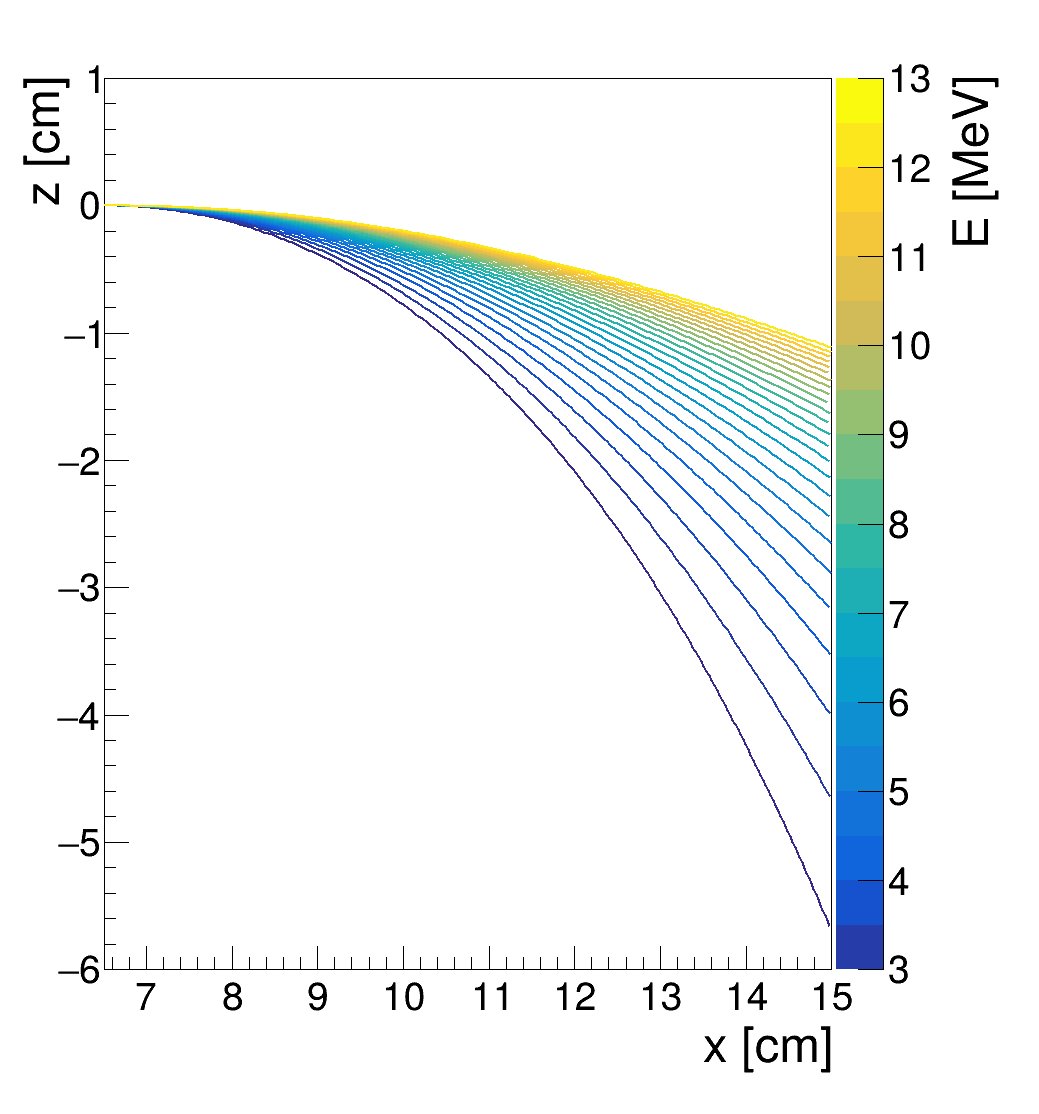}
    \caption{Set of 25 forward Runge--Kutta tracks in the OFTPC magnetic field, with initial position $(6.51,0,0)$\,\unit{\cm}, initial direction $(1,0,0)$, and kinetic energies from 3 to \qty{13}{\MeV} (colour scale), projected onto the $x$--$z$ plane. Higher-energy tracks are deflected less, and the trajectories of neighbouring energies bunch together as the energy increases; a fixed transverse displacement---for example from multiple scattering, which is not included in the simulation---therefore translates into a progressively larger energy error at higher energies.}
    \label{fig:rk_forward}
\end{figure}

To develop and test the reconstruction procedure, a simulation of the primary tracks is needed. The movement of primary particles, together with the production of ionization electrons in the primary vertices, was simulated using the HEED program (High Energy Electro-Dynamics)~\cite{SMIRNOV2005474}, which is part of the \garfieldpp toolkit. HEED does not apply multiple Coulomb scattering to the primary particle in the gas. Estimated from the Highland formula~\cite{Tanabashi:2018oca} over the primary path through the gas (up to about \qty{10}{\cm}), the RMS deflection is largest for low-energy electrons---approximately $10^\circ$ at \qty{1}{\MeV}, falling to about $1^\circ$ at \qty{13}{\MeV}. Although the deflection is therefore smallest at high energy, its effect on the reconstructed energy is significant across the whole range: as the energy increases, the trajectories of neighbouring energies become increasingly close (Fig.~\ref{fig:rk_forward}), so that a small transverse displacement translates into a large energy difference. Neglecting multiple scattering therefore makes the energy resolution quoted below an optimistic estimate, most notably at the upper end of the energy range.

\begin{figure}[H]
    \centering
    \begin{subfigure}{0.31\textwidth}
        \includegraphics[width=5cm]{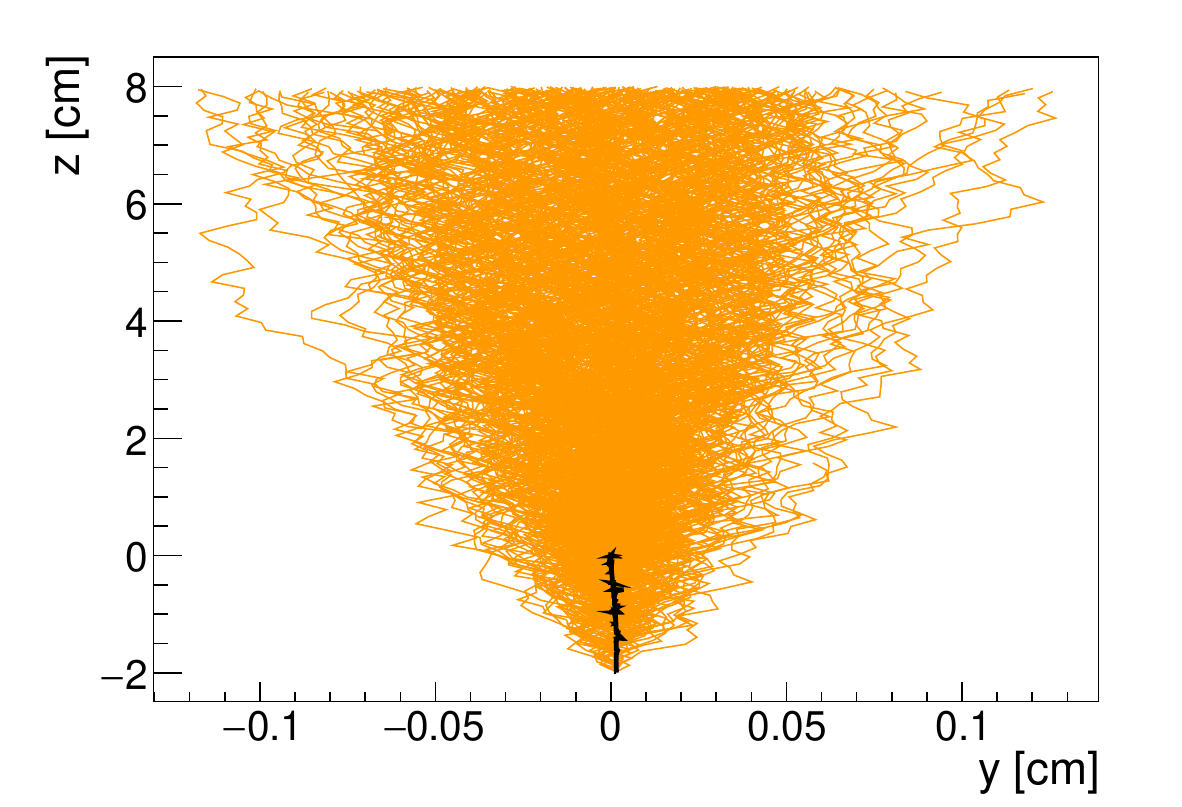}
        \caption{Diffusion \\ {\ \ \ \ \ \ front view }}
        \label{fig:TPC_diffusion_front_view}
    \end{subfigure}
    \hfill
    \begin{subfigure}{0.31\textwidth}
        \includegraphics[width=5cm]{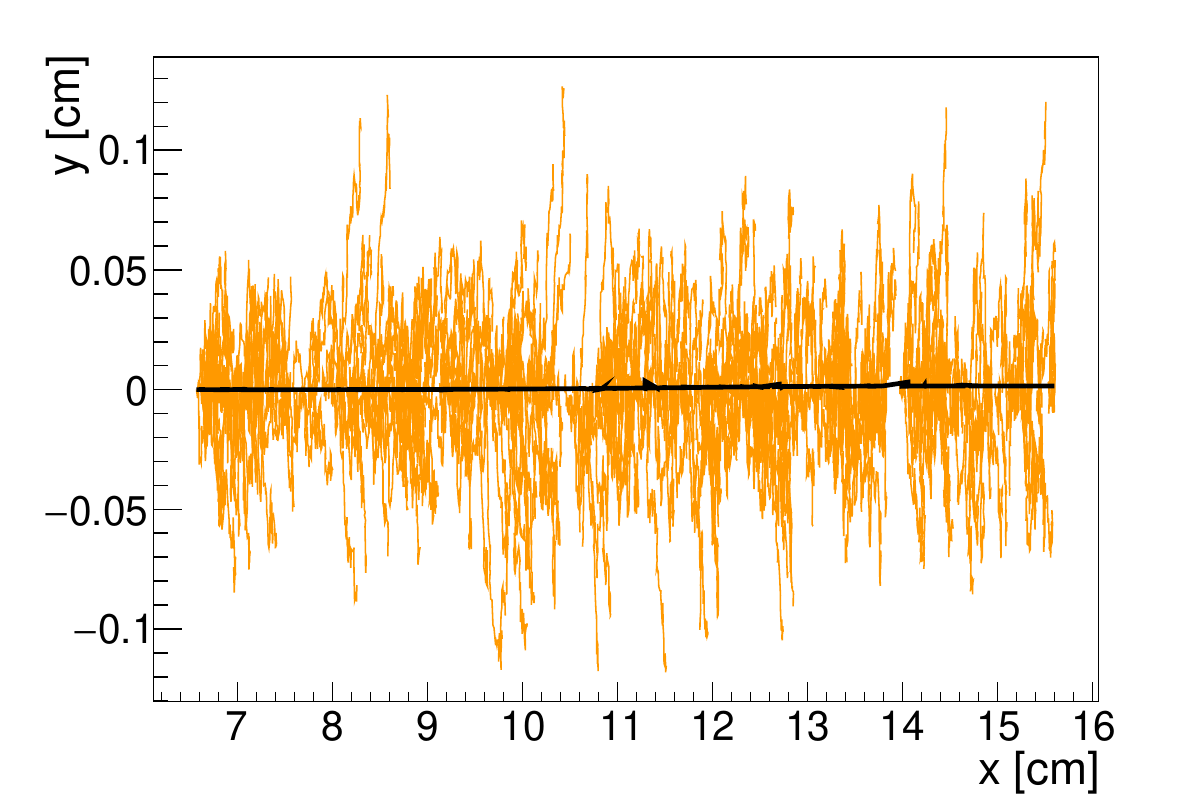}
        \caption{Diffusion  \\ {\ \ \ \ \ top view }}
        \label{fig:TPC_diffusion_top_view}
    \end{subfigure}
    \hfill
    \begin{subfigure}{0.31\textwidth}
        \includegraphics[width=5cm]{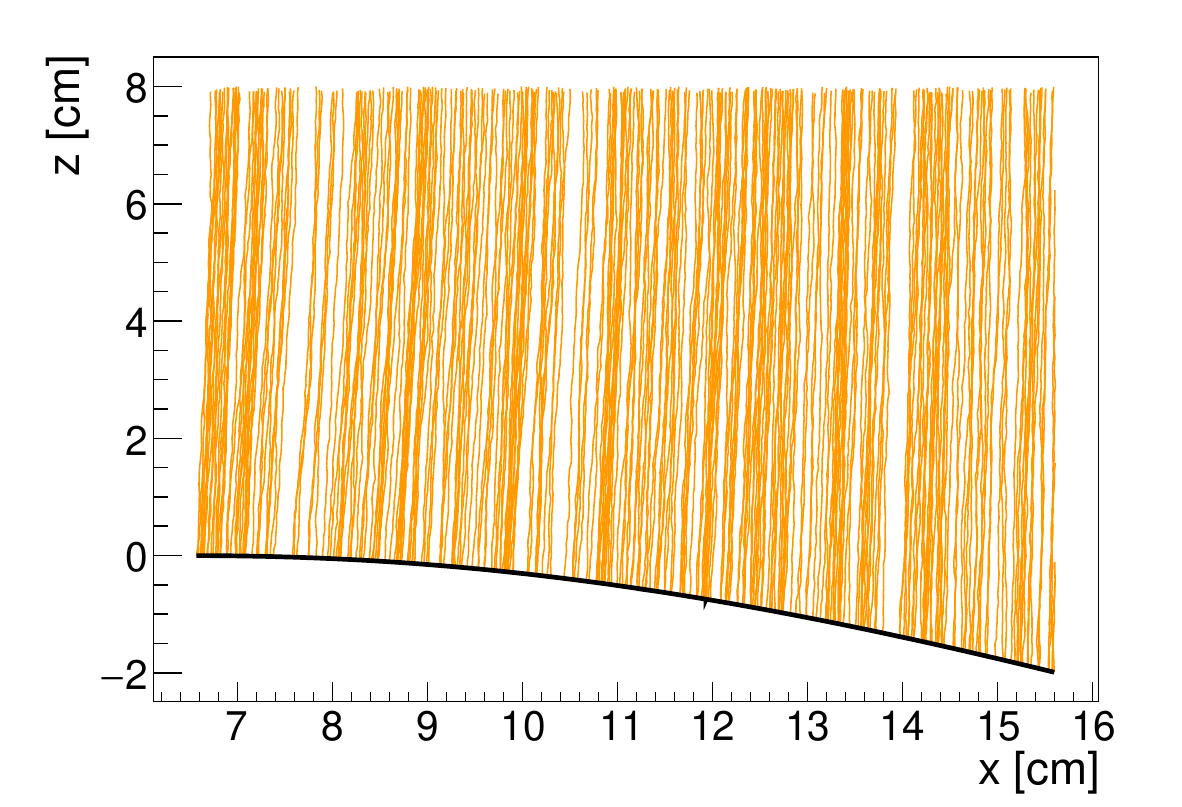}
        \caption{Drift of ionization electrons}
        \label{fig:TPC_drift_view}
    \end{subfigure}
    \caption{Simulation showing a track of a primary particle (drawn in black) and corresponding tracks of ionization electrons (drawn in orange). The projections shown in figures (a) and (b) demonstrate the randomness of the process, which is why the averaged direct mapping $\overline{\mathcal{M}}$ (Eqs.~\ref{average_mapping} and~\ref{inverse_mapping}) and its inverse $\overline{\mathcal{M}}^{-1}$ are used. 
    }
    \label{fig:TPC_diffusion_drift}
\end{figure}

The response of the OFTPC to primary tracks with different initial parameters, forming a regular grid that covers the kinematic phase space of tracks, was simulated on the MetaCentrum computer grid (see Acknowledgments). In the case of beryllium, the grid covers the kinematic range of both processes. The IPC pairs share the available transition energy ($\approx\qty{17}{\MeV}$ minus rest masses) with a broad, continuous partition between the two leptons, whereas the leptons from the two-body decay of the X17 ($m_{X17}\approx 17\,\mathrm{MeV}/c^2$) cluster near half the boson mass, i.e.\ around \qty{8}{\MeV} each. A grid spanning kinetic energies of \qtyrange{3}{13}{\MeV} therefore covers the region in which the X17 signal has to be separated from the IPC continuum. The grid was chosen to have 21 points in $\theta$ over $[-17.1^\circ, 17.1^\circ]$, 21 points in $\phi$ over $[-16.3^\circ, 16.3^\circ]$, and 11 points in $E_\mathrm{in}$ over $[3, 13]$\,MeV, for both electron and positron primaries. An example of a simulated track is shown in Fig.~\ref{fig:TPC_diffusion_drift}; the primary track is drawn in black, and the ionization electrons, drifting and diffusing towards the readout, in orange.

\begin{figure}[H]
	\centering
	\includegraphics[width=0.6\textwidth]{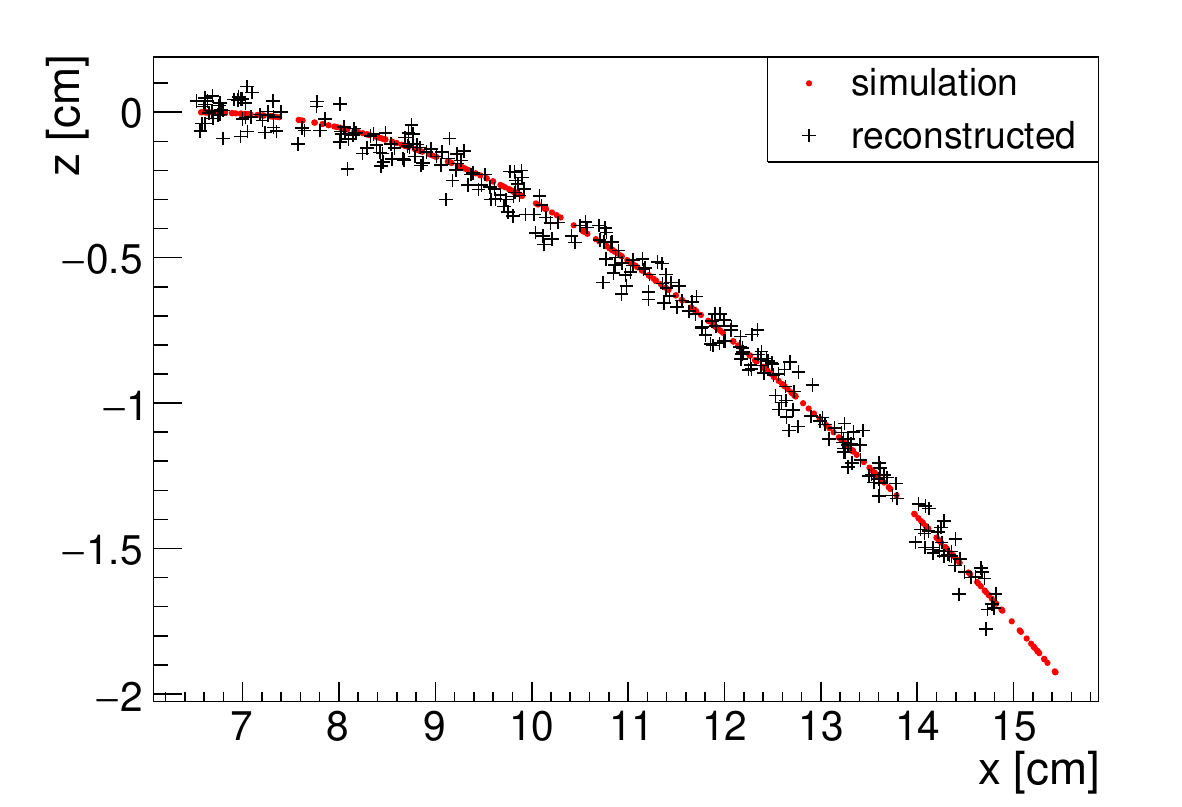}
	\caption{Simulated (in red color) and reconstructed interaction points for the simulated primary track with the kinetic energy of \qty{8}{\MeV} electron entering into the OFTPC at the position $(6.51,0,0)$\,\unit{\cm} (which corresponds to the center of the window of the OFTPC) and having direction vector $(1,0,0)$.}
	\label{fig:reco_track}
\end{figure}

To study the limitations of the reconstruction procedure, we first consider an idealized,  \emph{continuous} readout plane. For the ionization electrons produced by a~(simulated) primary track $(x_i, y_i, z_i)$ the corresponding reconstructed vertices $(\tilde{x}_i, \tilde{y}_i, \tilde{z}_i)$ are found by applying $\overline{\mathcal{M}}^{-1}$ to the readout positions, see the example in Fig.~\ref{fig:reco_track}. The $z$-coordinates of the vertices are reconstructed based on the time information, whose measurement starts when the primary track triggers the detector. Here, it happens when a particle enters the OFTPC volume; in reality, we expect it will be triggered by the MWPC detector. As it only takes about \qty{0.5}{\ns} from the creation of a primary particle to its exit from the entire detector, this difference is negligible, even compared to the fluctuation of the drift time.

\begin{figure}[H]
    \centering
    \includegraphics[width=0.48\textwidth]{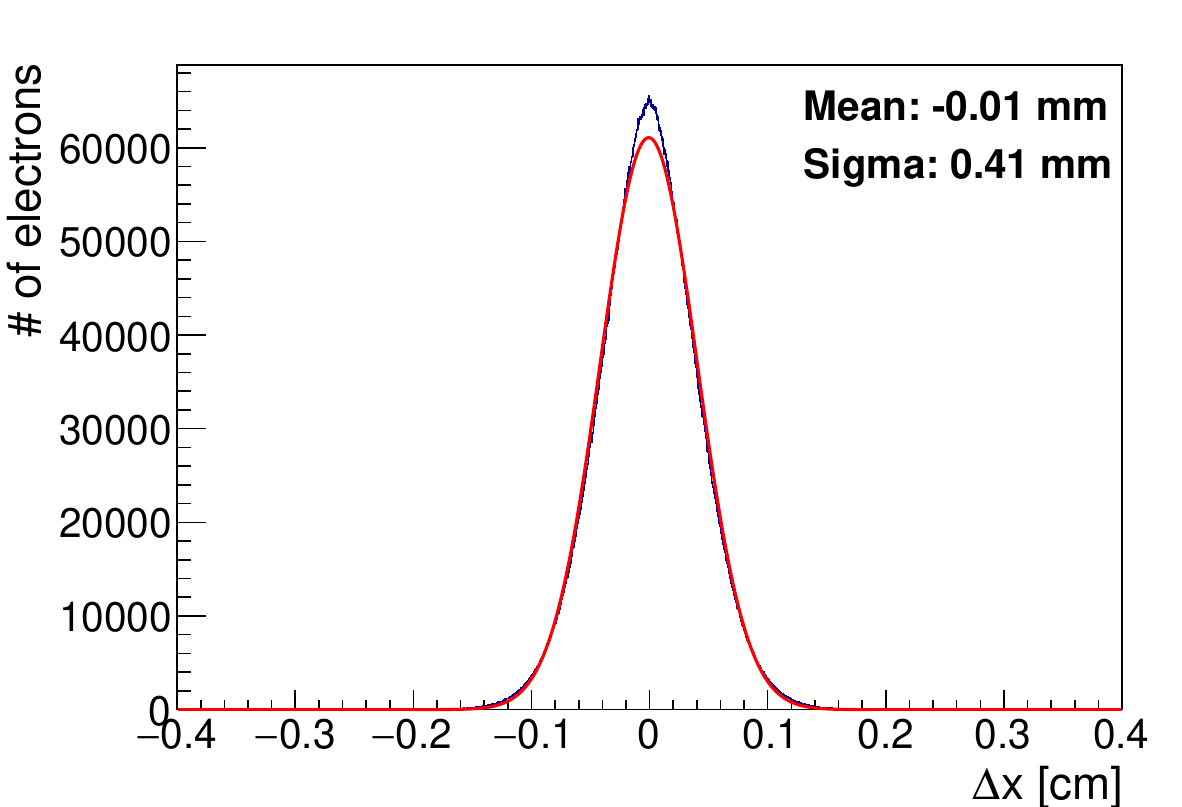}
    \hfill
    \includegraphics[width=0.48\textwidth]{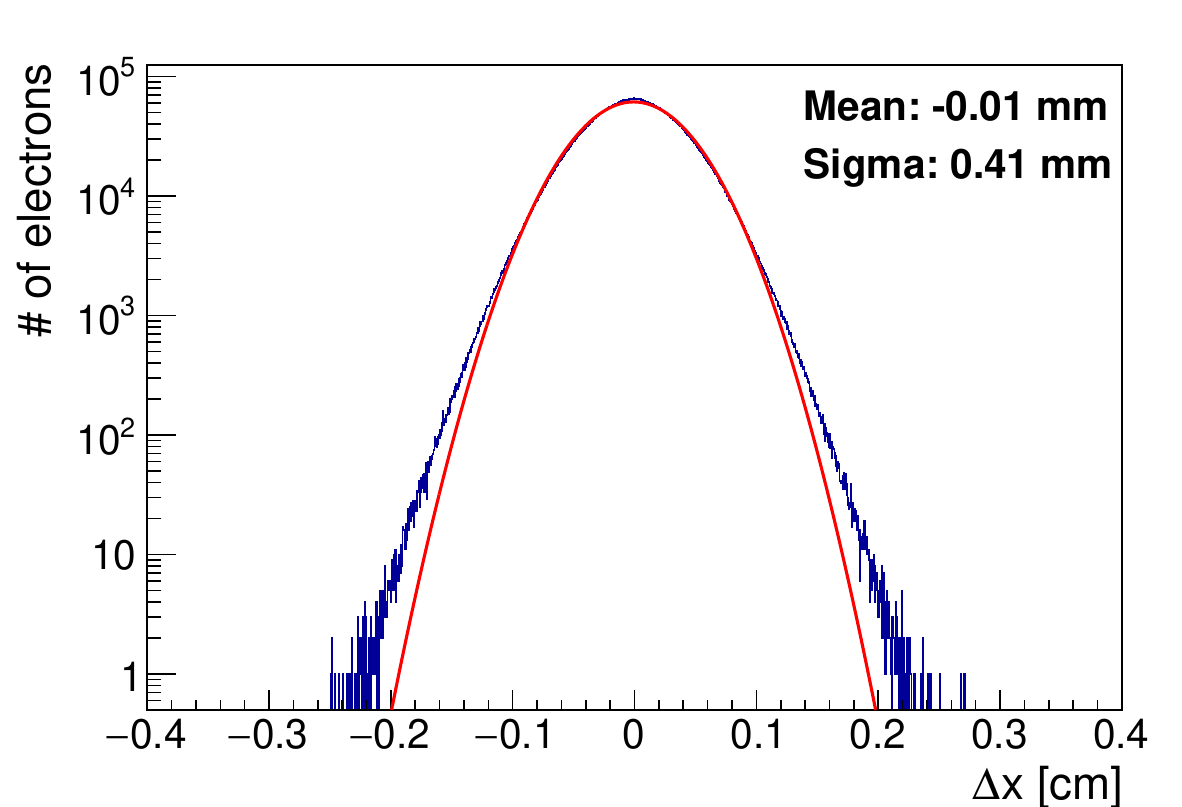}
    \includegraphics[width=0.48\textwidth]{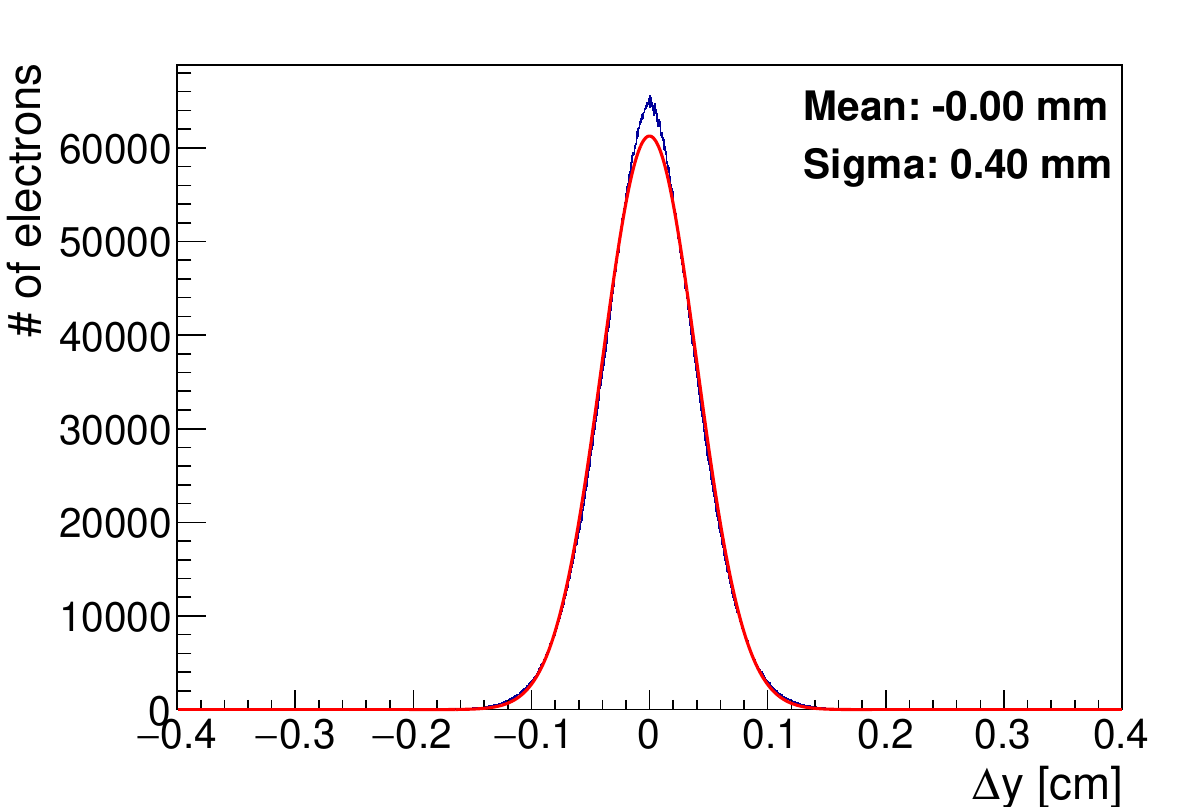}
    \hfill
    \includegraphics[width=0.48\textwidth]{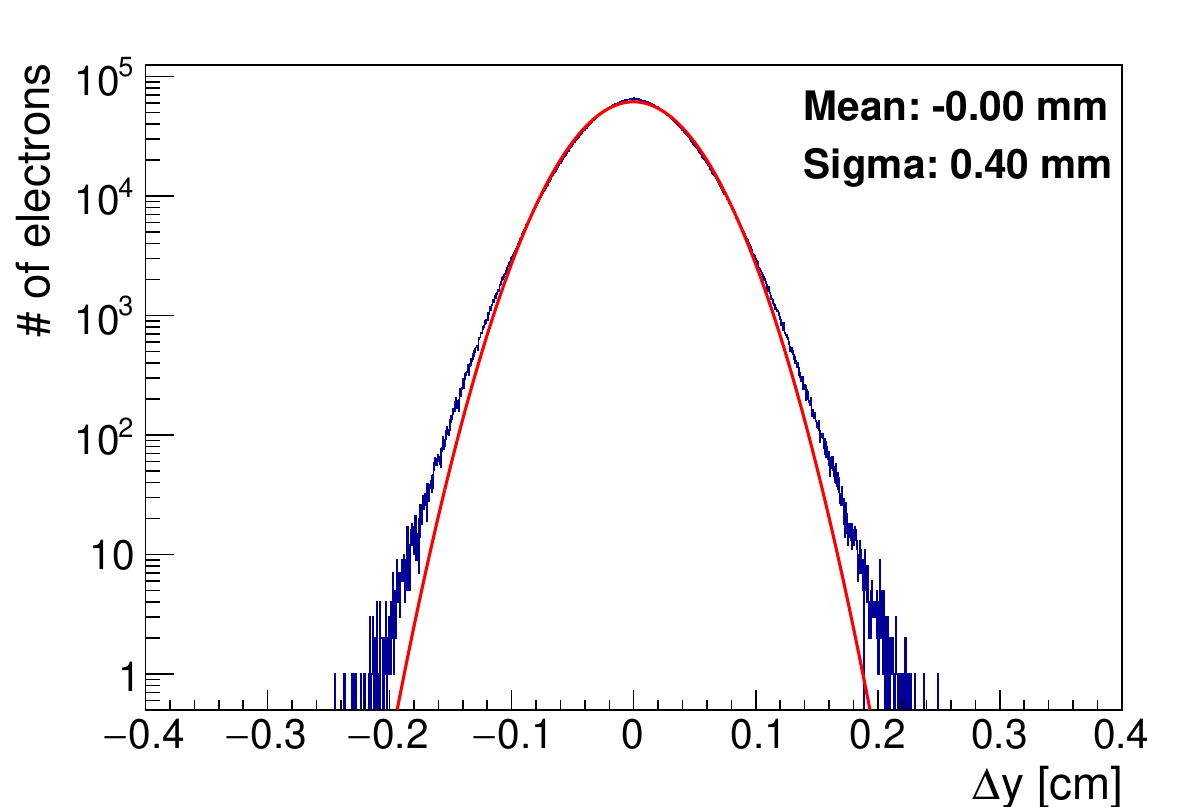}
    \includegraphics[width=0.48\textwidth]{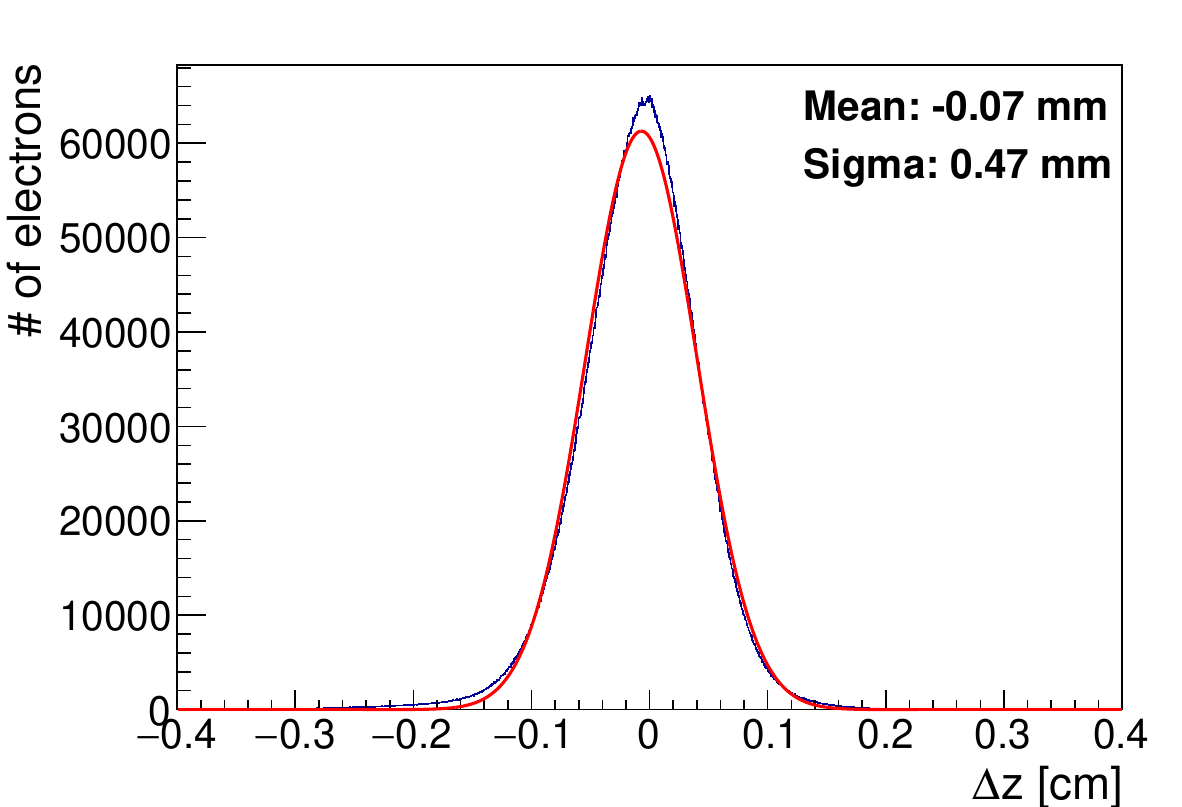}
    \hfill
    \includegraphics[width=0.48\textwidth]{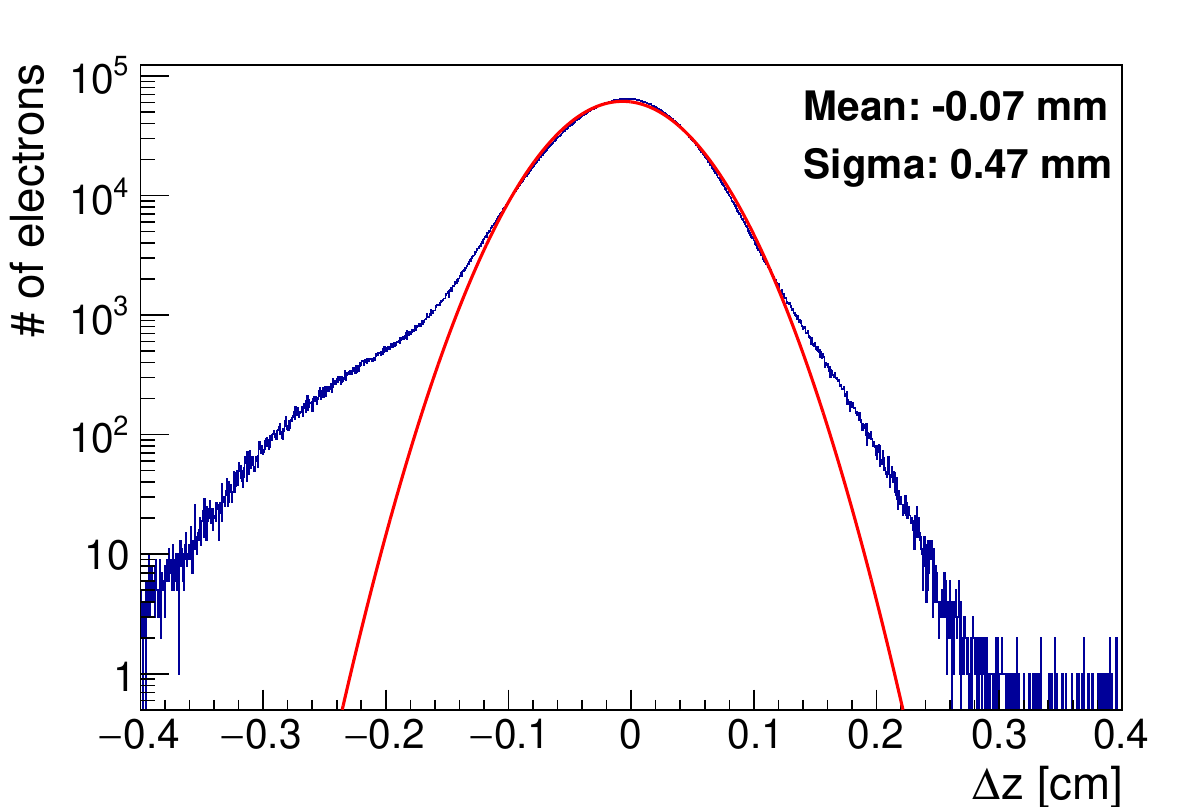}
    \caption{Gaussian fit of $x$, $y$, and $z$ residuals of reconstructed vertex coordinates, i.e., differences of original and reconstructed vertices $(x, y, z) - (\tilde{x}, \tilde{y}, \tilde{z})$. Histograms in linear (left) and logarithmic (right) scales are shown. Data from \num{14277175} electrons from \num{48485} simulated tracks were used. The statistical uncertainty of the original histograms is very small and is omitted. In terms of shape, the Gaussian fits do not perfectly capture the tails of the distribution; the dependence of the residuals on the track parameters will be studied further in future work.}
    \label{fig:residues_hist}
\end{figure}

The histograms of the residuals are summarized in Fig.~\ref{fig:residues_hist}. The errors are caused by the diffusion of the drifting ionization electrons resulting from their random collisions with gas particles (an effect that the averaged map cannot account for; see, for example, Fig.~\ref{fig:TPC_diffusion_drift}). We also see a small systematic shift in $z$, whose origin is still under investigation; it may be related to the treatment of the non-zero initial velocities of the ionization electrons.

\begin{figure}[H]
	\centering
	\includegraphics[width=0.85\textwidth]{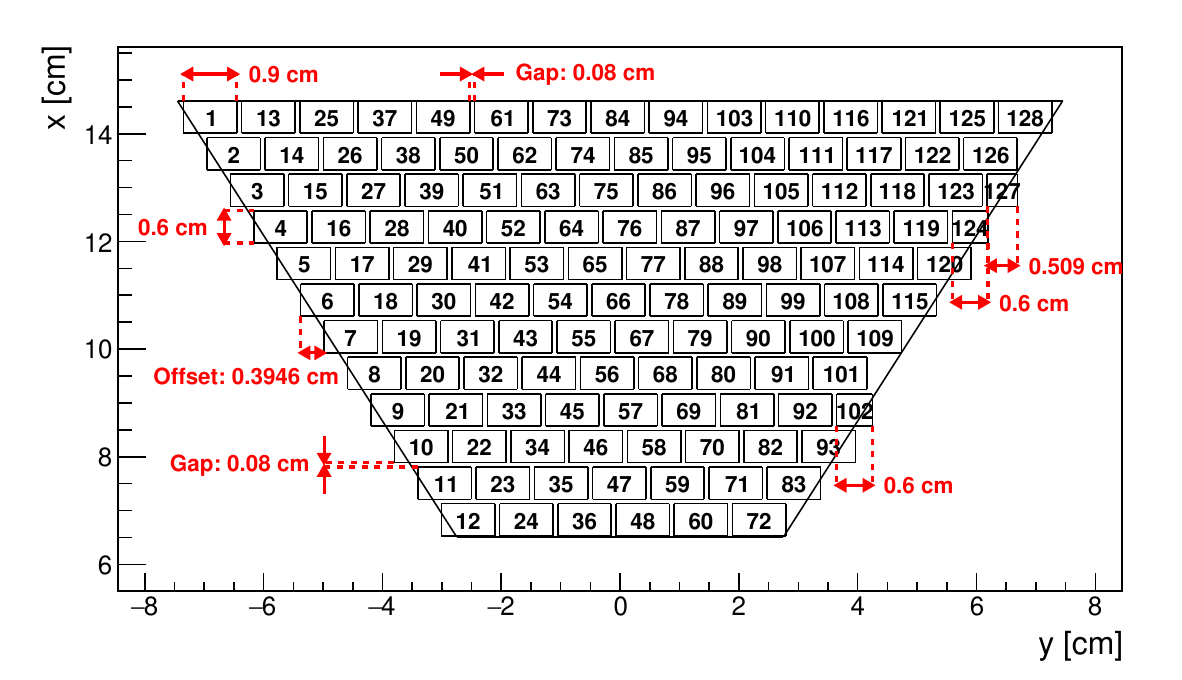}
    \vspace{-0.5cm}
	\caption{Pad layout of one of the OFTPC readout planes and its parameters. Pads 102, 124, and 127 are irregular; the rest have the same dimensions.}
	\label{fig:pads}
\end{figure}

In reality, the readout plane is divided into pads, as shown in Fig.~\ref{fig:pads}. The charge collected by each pad is assigned to its center and to the center of the time bin, which yields a sparser sampling of the inverse map \(\overline{\mathcal{M}}^{-1}\) used for track reconstruction. In the future, this approach might be refined using a~likelihood method. We can precalculate the map for these points if the reconstruction speed becomes an issue.

The multiplying effect of the GEM foil is not accounted for\footnote{The GEMs might affect the resolution of energy reconstruction due to a) their gain fluctuation (which can lower the energy resolution) and b) charge spreading/sharing (which can improve the energy resolution).}, and for the reconstruction procedure, all pads are considered to have the same area, although there are three special pads near the detector's edge. Using the inverse map $\overline{\mathcal{M}}^{-1}$ at the center of each hit pad and time bin, the reconstructed voxels of the primary track are found, and a weight equal to the number of ionization electrons is assigned to each of them.

\subsection{Energy reconstruction}
The track passing through the reconstructed vertices is determined by the MIGRAD minimization algorithm implemented in ROOT \cite{Brun:1997pa}. The minimized function is the standard weighted sum of squared distances between the reconstructed positions of the primary vertices and the trajectory predicted for a given value of the single free parameter, the particle energy. The predicted trajectory is calculated with the Runge--Kutta algorithm in the inhomogeneous magnetic field of the OFTPC, using as input (a) the direction vector of the track (spherical angles $\theta$ and $\varphi$), (b) the track origin in the TPC window $(y_0,z_0)$, and (c) the description of the inhomogeneous magnetic field within the OFTPC volume, via a grid of points.

\begin{figure}[H]
	\centering
	\begin{subfigure}[t]{0.48\textwidth}
		\includegraphics[width=7cm]{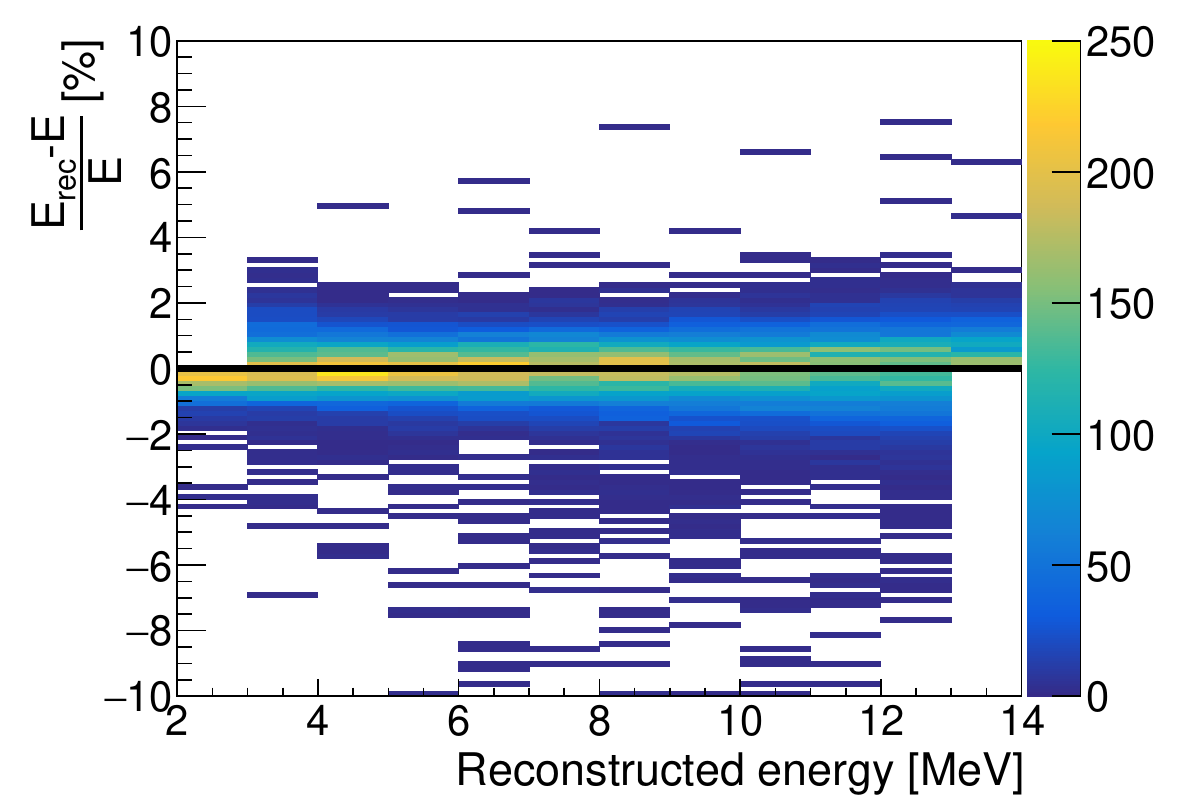}
		\caption{Reconstructed energy accuracy as a function of reconstructed energy \(E_{\mathrm{rec}}\) for primary electrons.}
		\label{fig:delta_energy_energy_e}
	\end{subfigure}
	\hfill
	\begin{subfigure}[t]{0.48\textwidth}
	\includegraphics[width=7cm]{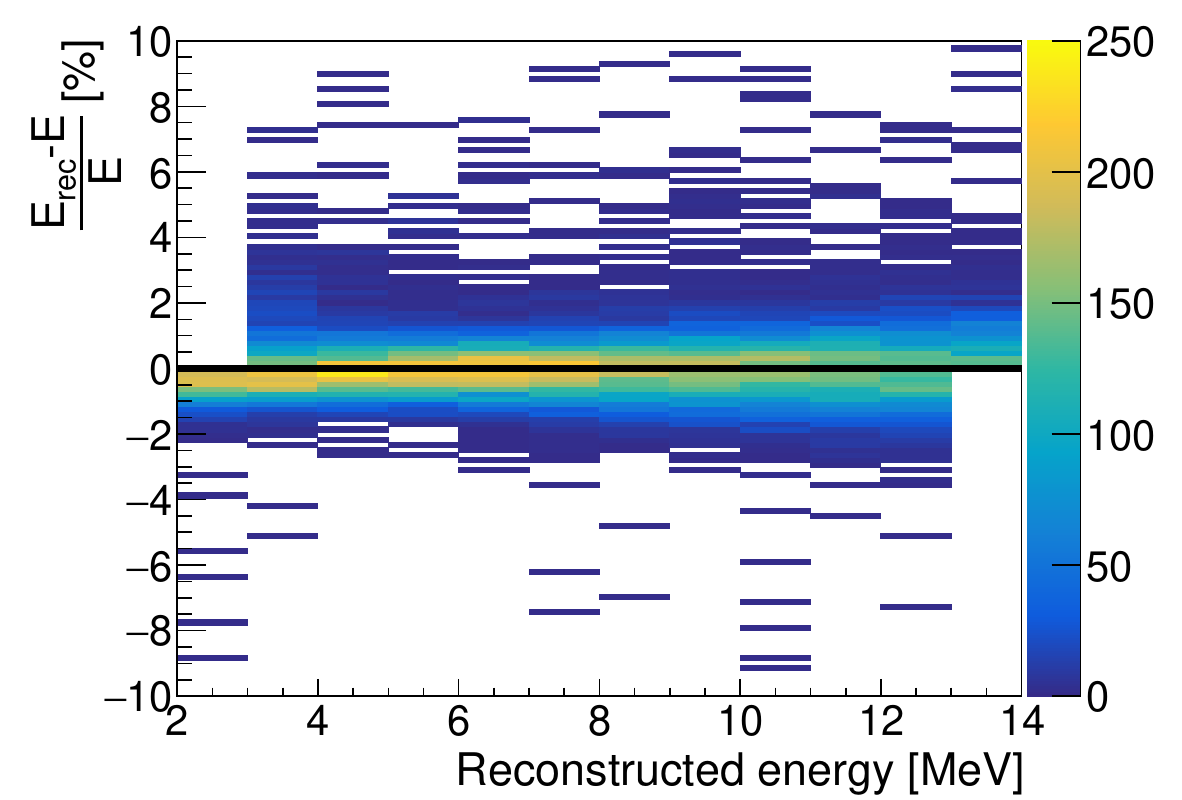}
		\caption{Reconstructed energy accuracy as a function of reconstructed energy \(E_{\mathrm{rec}}\) for primary positrons.}
		\label{fig:fig:delta_energy_energy_p}
	\end{subfigure}
	\bigskip
	\begin{subfigure}[t]{0.48\textwidth}
		\includegraphics[width=7cm]{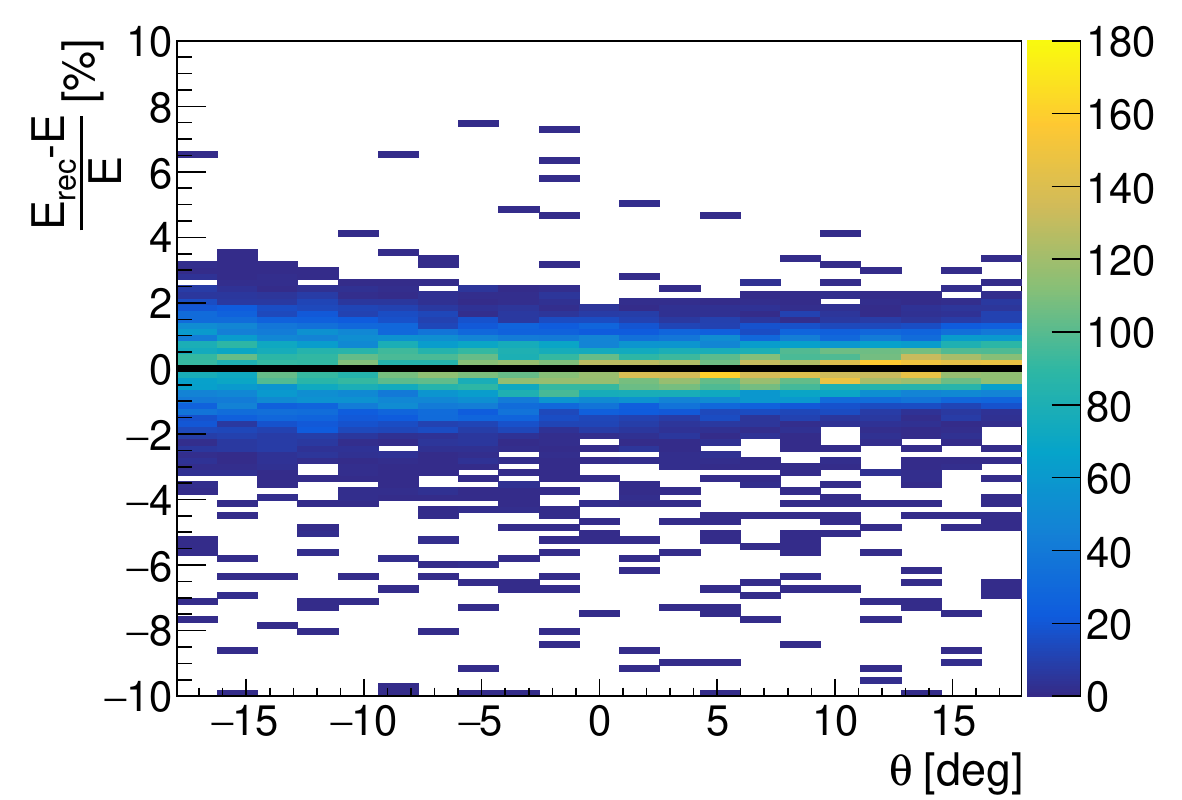}
		\caption{Reconstructed energy accuracy as a function of simulated polar angle \(\theta\) for primary electrons.}
		\label{fig:delta_energy_theta_e}
	\end{subfigure}
        \hfill
	\begin{subfigure}[t]{0.48\textwidth}
		\includegraphics[width=7cm]{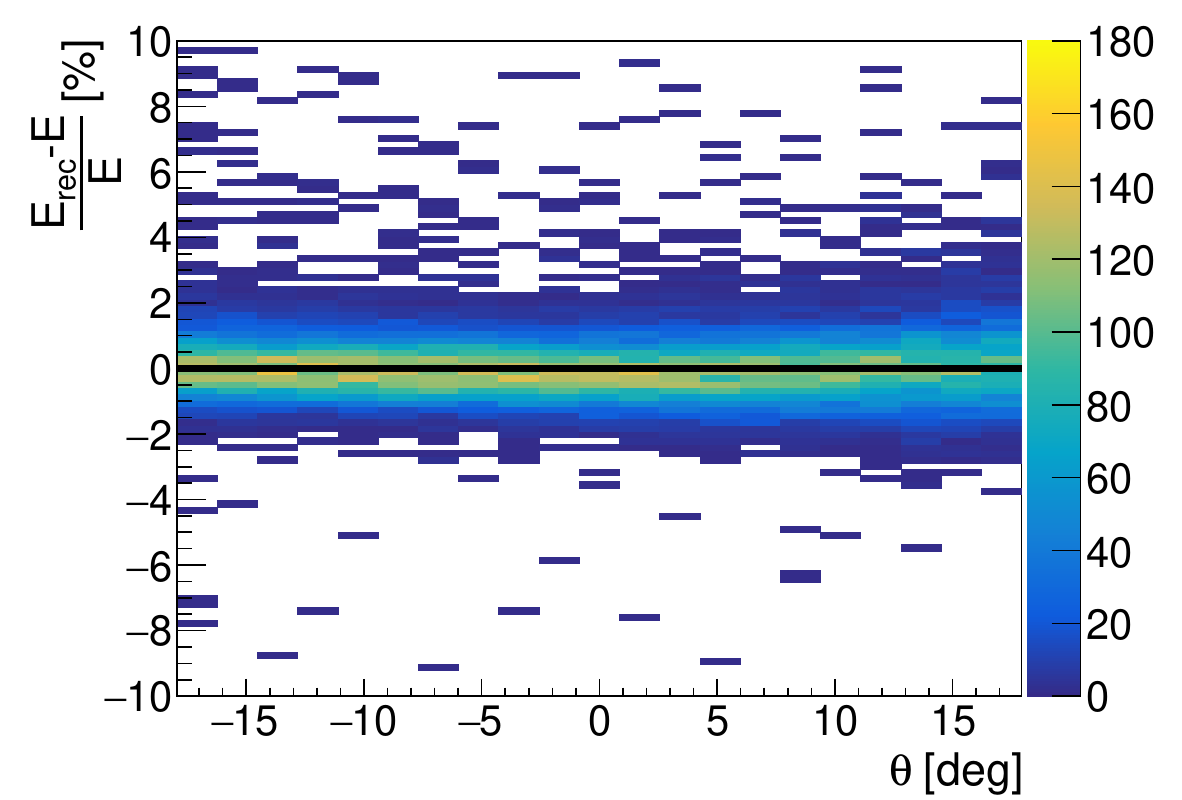}
		\caption{Reconstructed energy accuracy as a function of simulated polar angle \(\theta\) for primary positrons.}
		\label{fig:delta_energy_theta_p}
	\end{subfigure}
 	\bigskip
	\begin{subfigure}[t]{0.48\textwidth}
		\includegraphics[width=7cm]{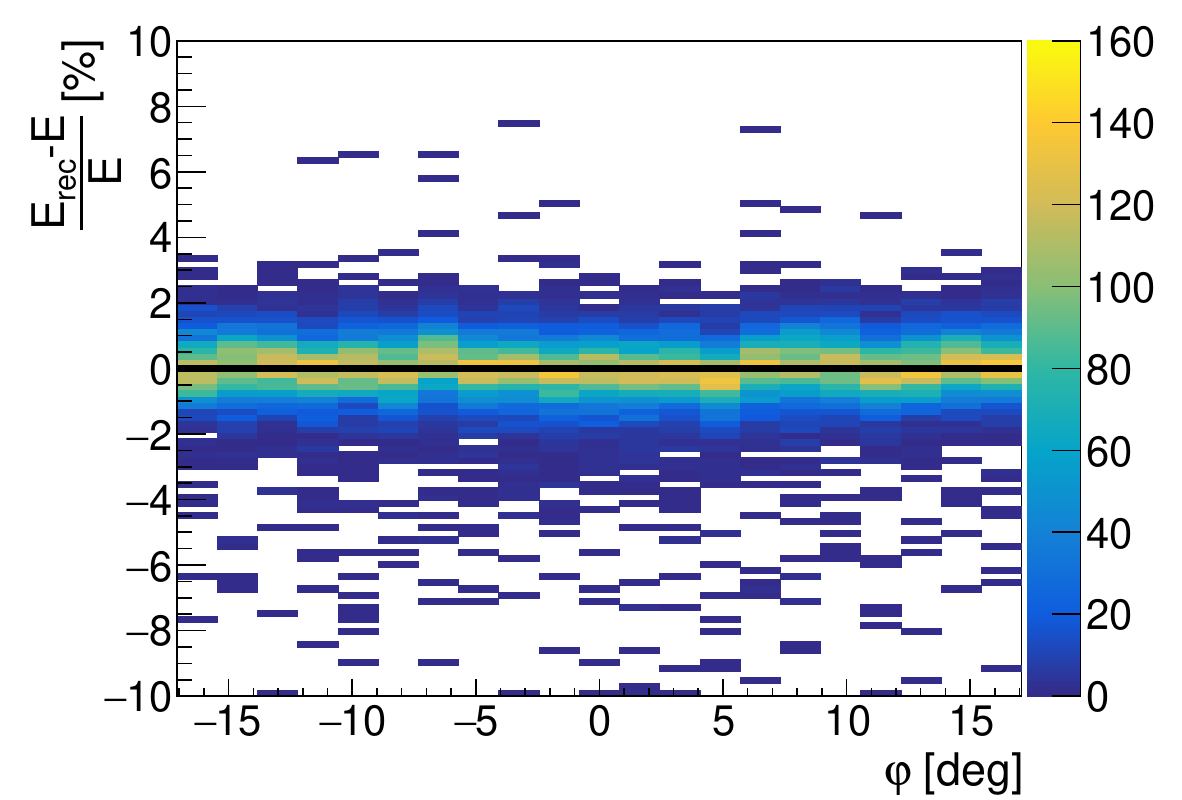}
		\caption{Reconstructed energy accuracy as a function of simulated  azimuthal angle \(\varphi\) for primary electrons.}
		\label{fig:delta_energy_phi_e}
	\end{subfigure}
	\hfill
        \begin{subfigure}[t]{0.48\textwidth}
		\includegraphics[width=7cm]{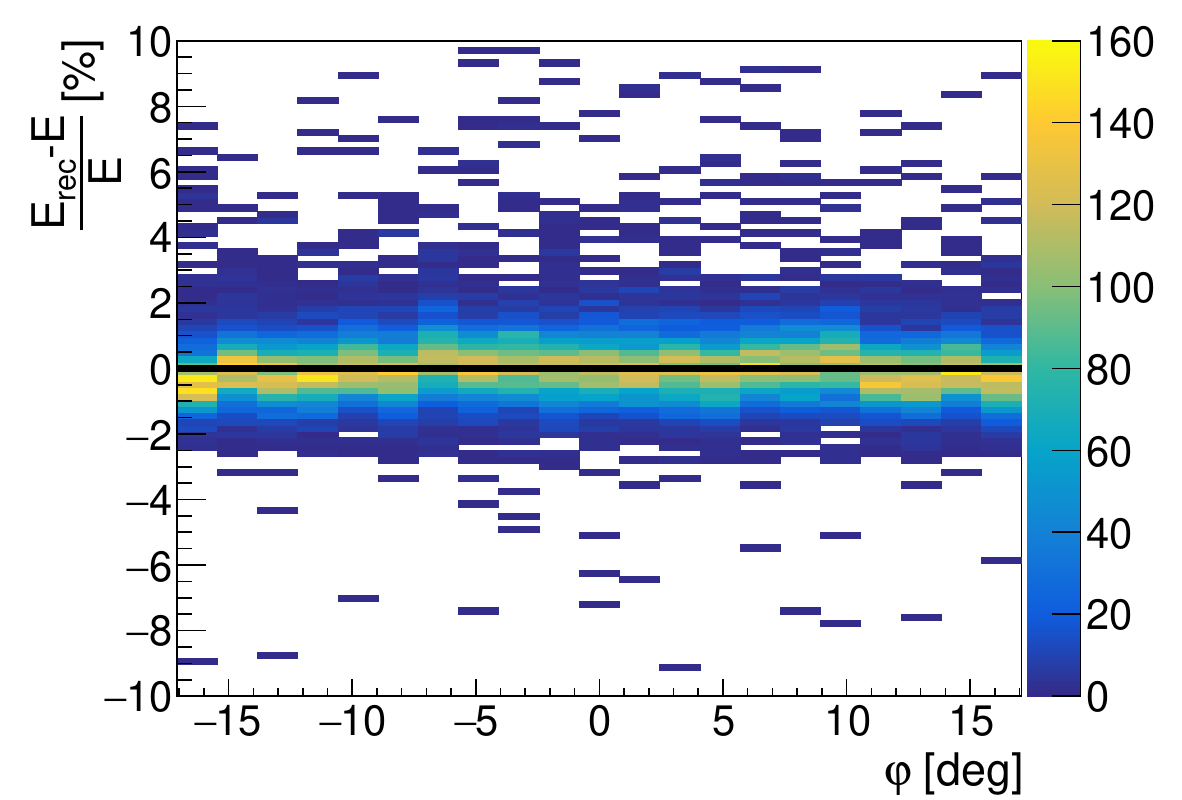}
		\caption{Reconstructed energy accuracy as a function of simulated azimuthal angle \(\varphi\) for primary positrons.}
		\label{fig:delta_energy_phi_p}
	\end{subfigure}
	\caption{Accuracy of reconstructed energy, both for electrons (in the left column) and positrons (in the right column) as a function of reconstructed energy (the upper line), polar angle \(\theta\) (the middle line), and azimuthal angle \(\varphi\) (the bottom line).}
	\label{fig:delta_energy}
\end{figure}

\begin{figure}[H]
	\centering
	\begin{subfigure}{\textwidth}
		\includegraphics[width=7.4cm]{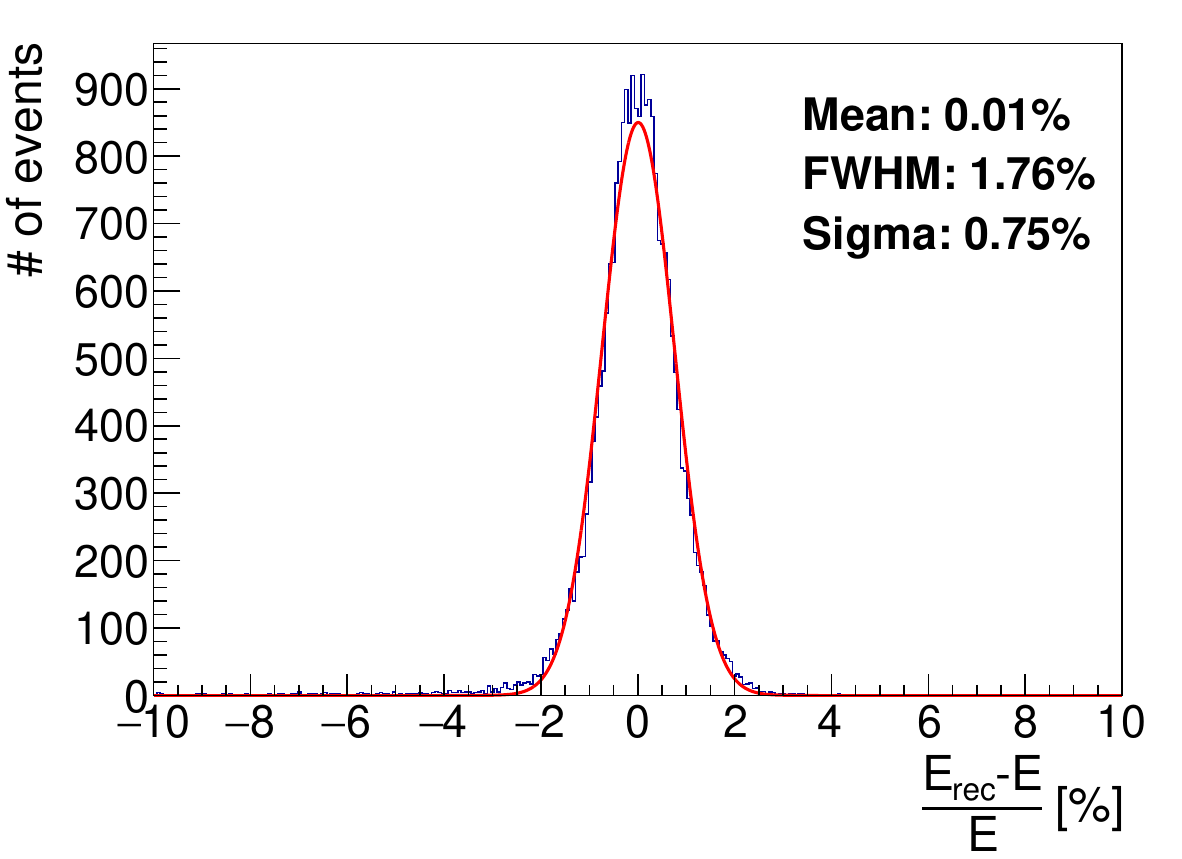}
        \hfill
		\includegraphics[width=7.4cm]{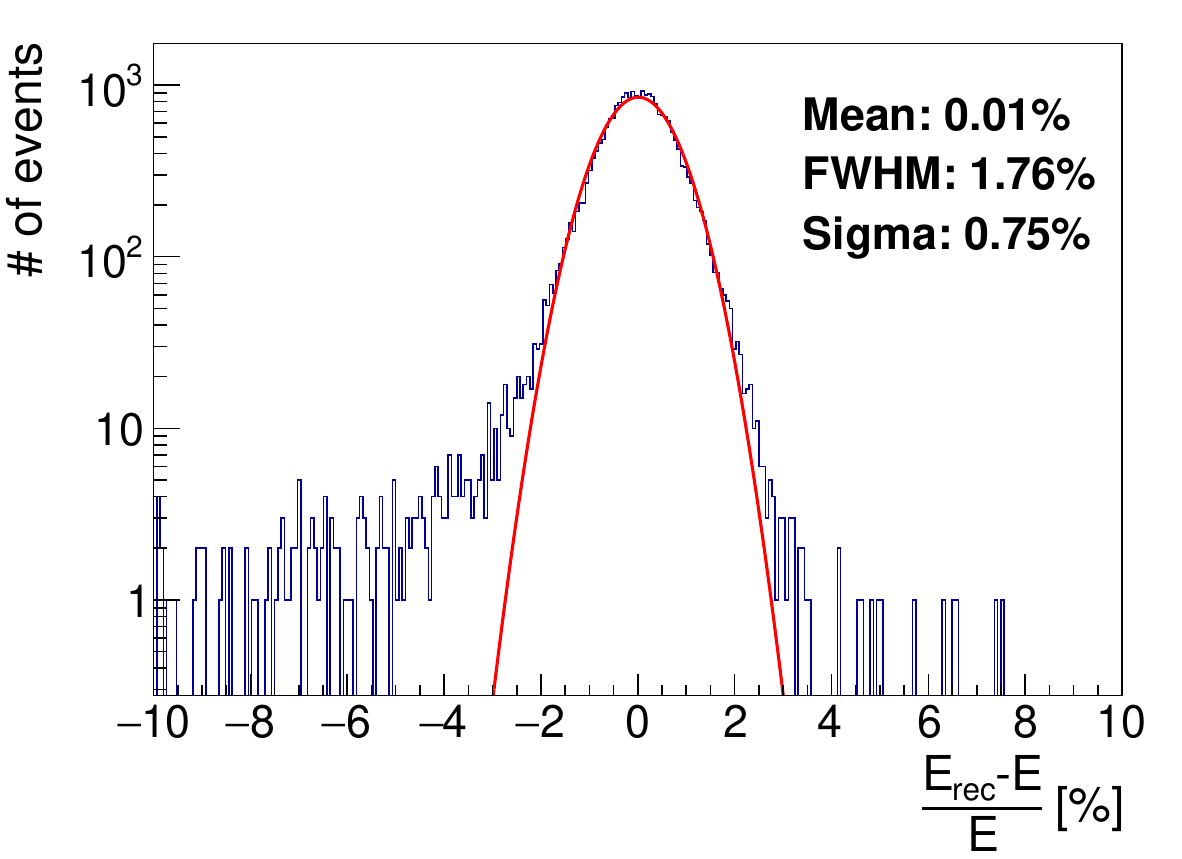}
		\caption{Energy reconstruction accuracy for electrons using linear (left) and logarithmic (right) scale.}
		\label{fig:TPC_energy_reco_err_e}
	\end{subfigure}
	\hfill
	\begin{subfigure}{\textwidth}
		\includegraphics[width=7.4cm]{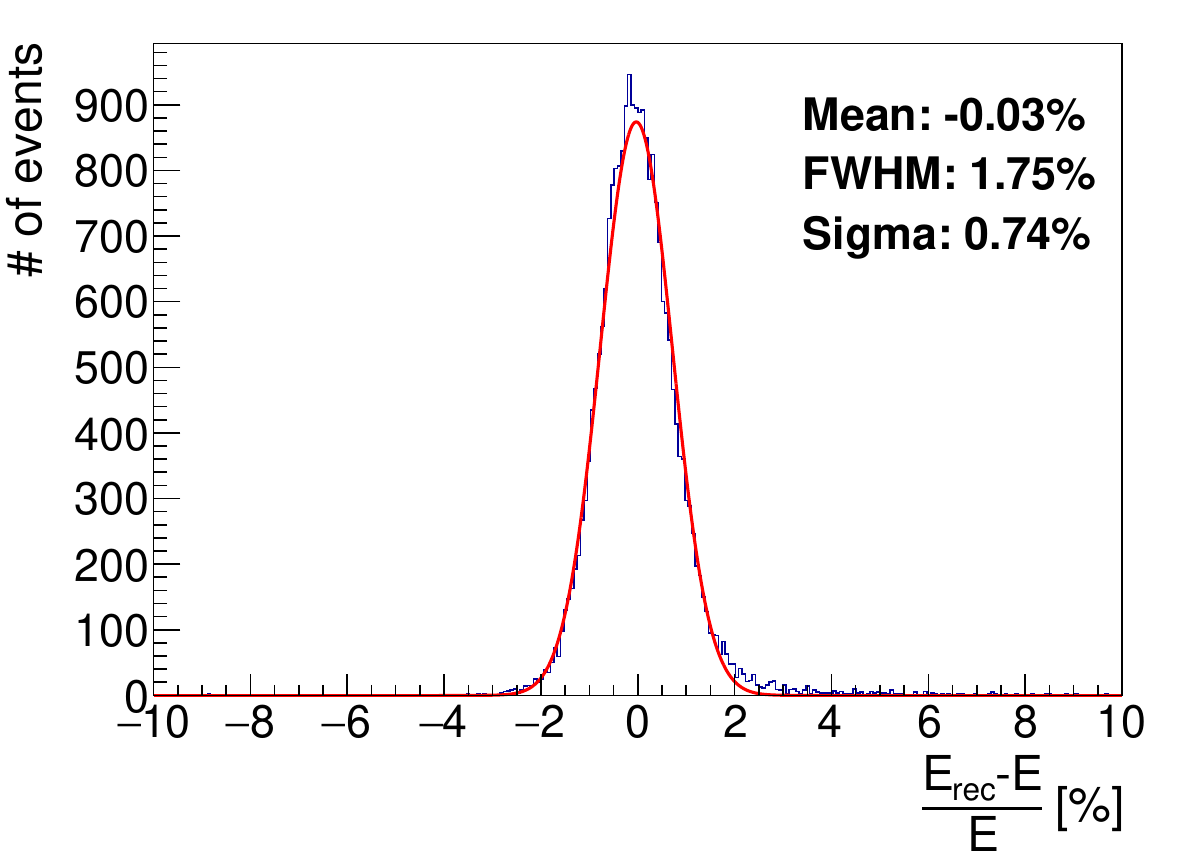}
        \hfill
		\includegraphics[width=7.4cm]{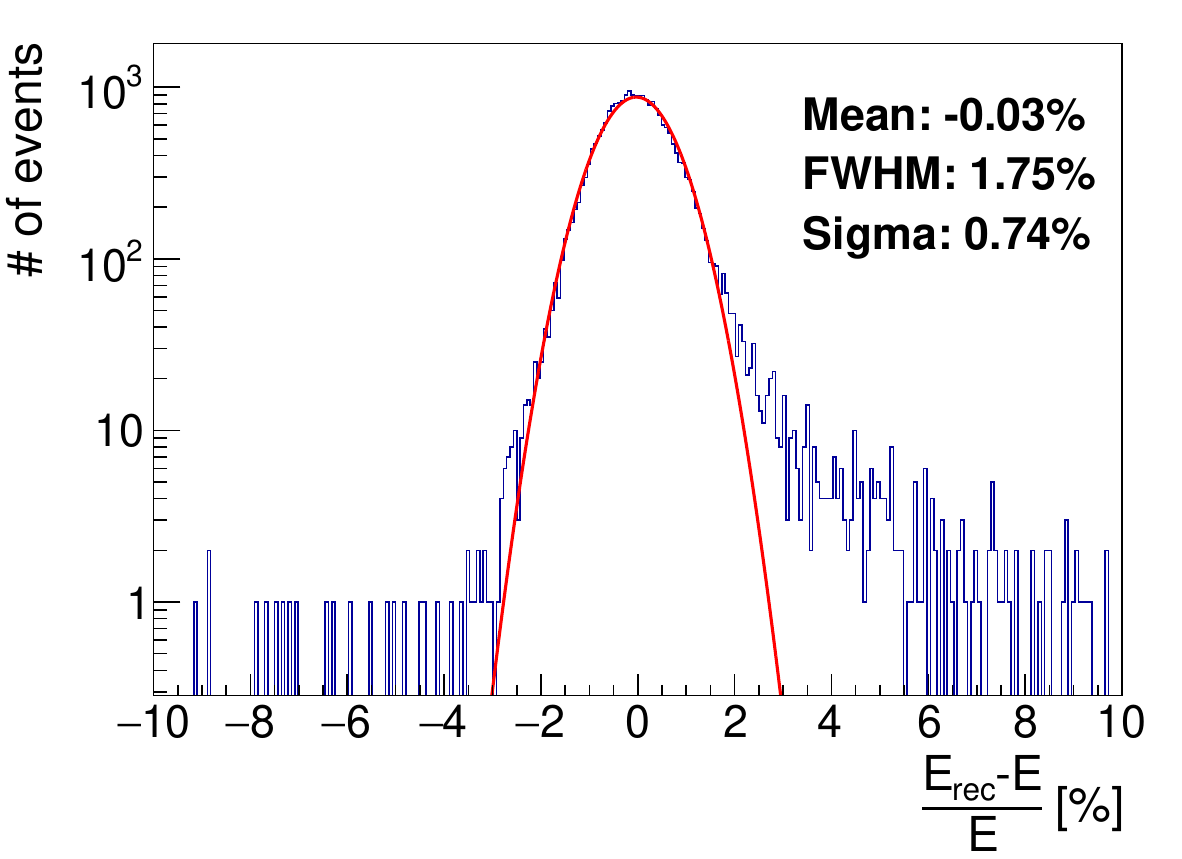}
		\caption{Energy reconstruction accuracy for positrons using linear (left) and logarithmic (right) scale.}
		\label{fig:TPC_energy_reco_err_p}
	\end{subfigure}
	\caption{Energy reconstruction accuracy. The width is quantified by the sigma of a Gaussian fit to the central part of the $\Delta E / E$ distribution. The Gaussian sigma (below 1\%) and the corresponding FWHM ($\approx 1.8\%$, i.e.\ $2.355\,\sigma$ for the fitted core) are both quoted in the Discussion.}
	\label{fig:TPC_energy_reco_err}
\end{figure}

The results are summarized as functions of track parameters: reconstructed energy $E_{\mathrm{rec}}$, polar $\theta$ and azimuthal $\varphi$ angles, for the segmented readout plane, in Fig.~\ref{fig:delta_energy}. Systematic errors were corrected using a~3D linear fit in the three variables $E_\text{rec}$, $\theta$, and $\varphi$---applied separately for electrons and positrons---as an additive correction to $E_\text{rec}$ across the full simulated phase space.
The result of reconstructed energy resolution, without details about its energy and angular dependencies, is plotted in Fig.~\ref{fig:TPC_energy_reco_err}, and shows that the fitted Gaussian sigma of the relative energy reconstruction is better than 1\% for both electrons and positrons, with a FWHM of approximately 1.8\%.

\section{Discussion and conclusion}
\begin{figure}[H]
    \centering
    \includegraphics[width=0.55\linewidth]{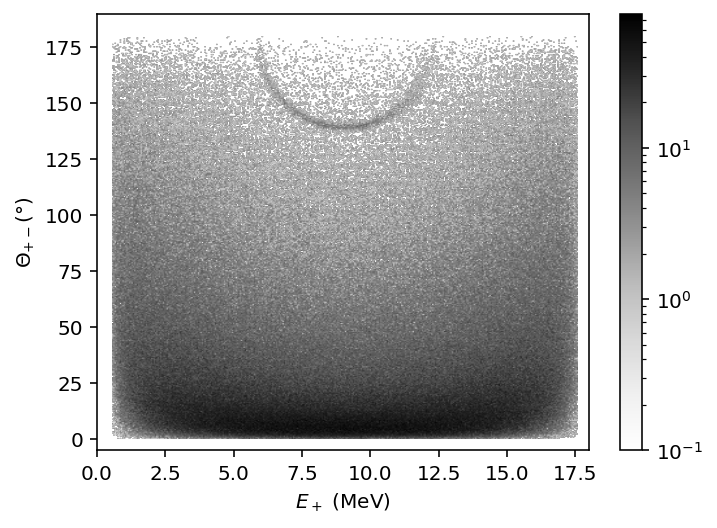}
    \caption{2D histogram of the $\pair$ opening angle vs.\ positron energy, as expected from a measurement. The data originate from a simulation folded with the dominant components of the detector response---the energy resolution (Fig.~\ref{fig:TPC_energy_reco_err}) and the angular resolution from~\cite{triangle}---and use $10^6$ events with a branching ratio X17/IPC = \num{3E-3}. The pattern of X17 events on top of the IPC background, while less sharp than in the ideal case of Fig.~\ref{fig:theta-E-plot}, remains clearly visible.}
    \label{fig:EvsTheta_real}
\end{figure}

By combining the energy-reconstruction precision (Fig.~\ref{fig:TPC_energy_reco_err}) and the angular resolution from~\cite{triangle}, the idealized case of Fig.~\ref{fig:theta-E-plot} becomes the partially realistic distribution shown in Fig.~\ref{fig:EvsTheta_real}, in which the dominant resolution effects are folded in. Even after accounting for this smearing, the X17 signal region remains clearly visible on top of the IPC background.

The achieved energy-reconstruction precision---Gaussian sigma below 1\% and FWHM of approximately 1.8\% for both electrons and positrons---is encouraging for a TPC operating with a strongly inhomogeneous magnetic field. These numbers were obtained under idealized simulation conditions (no electronic noise, no GEM amplification, exact knowledge of the magnetic field map and of the initial track direction), and several effects not yet accounted for will influence the final performance:
\begin{enumerate}[label=\alph*)]
    \item The real magnetic-field map may differ from the finite-element-method map used in the reconstruction. This effect can be mitigated by direct measurement of the field via a dedicated probe or by fitting straight laser or muon tracks.
    \item The direction vectors of the primary tracks are known only with finite precision, set by the combination of the TPX3 and MWPC measurements. Ultimately, a global fit---incorporating these parameters into the OFTPC fit---might be the best solution.
    \item The GEM amplification introduces gain fluctuations (broadening the charge distribution) and charge spreading (improving position resolution); the net effect on energy reconstruction has not yet been quantified.
    \item Multiple Coulomb scattering of the primary lepton in the gas is not included in the present simulation: native HEED treats the primary as a rigid straight-line track and applies multiple scattering only to the secondary $\delta$-electrons it produces~\cite{Garfield++,SMIRNOV2005474}. As discussed in Section~\ref{sec:simulation}, the RMS deflection over the OFTPC depth is small, but because the trajectories of neighbouring energies are closely spaced---increasingly so at high energy---it can degrade the reconstructed energy across the whole range. This is a known limitation of standalone HEED rather than of the reconstruction method itself: primary multiple scattering can be incorporated through a coupled Geant4--\garfieldpp interface, in which Geant4 transports the primary with its multiple-scattering models (Urban or Goudsmit--Saunderson) while HEED retains the microscopic generation and transport of the $\delta$-electrons~\cite{Pfeiffer2019}. Quantifying the effect with such a simulation is left for future work.
\end{enumerate}

In summary, the simulation demonstrates that the OFTPC reconstruction pipeline achieves sub-percent energy resolution under idealized conditions, and that the X17 signal region remains well visible on top of the IPC background after folding in the currently modeled detector effects (the energy and angular resolutions). This figure should be read as an optimistic, proof-of-principle estimate: effects not yet included---most notably multiple scattering of the primary in the gas, together with the magnetic-field-map fidelity and the trigger-timing precision---will degrade it, and quantifying them is the natural next step. The latter two are both addressable, the former through direct field mapping and the latter through careful MWPC calibration.

\section*{CRediT author statement}
\textbf{Martin Vav\v r\'ik}: Methodology, Software, Analysis, Investigation, Visualization, Writing. 
\textbf{Babar Ali}: Writing - Review \& Editing. 
\textbf{Hugo Natal da Luz}: Conceptualization, Investigation, Software, Formal Analysis, Validation,
	Supervision, Project Administration.
\textbf{Olivier Rousselle}: Software, Analysis, Investigation, Visualization.
\textbf{Rudolf S\'ykora}: Investigation, Review \& Editing.
\textbf{Tom\'a\v s S\'ykora}: Conceptualization, Validation, Investigation, Supervision, Writing including Original Draft Preparation.

\section*{Acknowledgments}
This work was supported by the GA\v CR - Czech Science Foundation
grant GA21-21801S.
Computational resources, MetaCentrum, were provided by the e-INFRA CZ project
(ID:90140), supported by the Ministry of Education, Youth and Sports
of the Czech Republic (MEYS).
The measurements with the proton beam took place in the Van de Graaff accelerator laboratory, supported by the LM2018108 grant of MEYS.
Figure~\ref{fig:OFTPC_coordinates} was prepared with GeoGebra\textsuperscript{\textregistered} under a non-commercial licence.
This article is intended to be published as open access. CTU Prague participates in the CzechELib Read \& Publish agreement with Elsevier, under which accepted articles by corresponding authors affiliated with CTU Prague are published open access under the Creative Commons Attribution 4.0 International (CC~BY~4.0) licence at no direct cost to the authors.

\bibliographystyle{ieeetr}
{\footnotesize
	\bibliography{TPCOF_reco_biblio}}

\end{document}